\documentclass[prb,twocolumn,floatfix]{revtex4-2}
\usepackage{latexsym}
\usepackage{graphicx}

\usepackage{subfigure}
\usepackage{array}
\usepackage{verbatim}
\usepackage{amsmath}
\usepackage{color}
\usepackage{xcolor}
\usepackage{soul}
\usepackage{tabu}
\usepackage{multirow}
\usepackage{lipsum}
\usepackage[normalem]{ulem}
\begin{document}
\graphicspath{{fig/}{./}}
\title{Excitonic-trion population in two-dimensional
  halide perovskites}
\author{Efstratios Manousakis}
  \address{
    Department  of  Physics, Florida State University,
    Tallahassee, Florida, 32306-4350\\
  and Department of Physics, Harvard University, Cambridge, MA, 02138\\
  and Department of Physics, National and Kapodistrian University of Athens,
  157 84 Athens, Greece}
\date{\today}
\begin{abstract}
  There are many reports of a surprisingly high charge-carrier density
  with sizable mobility in photo-excited two-dimensional (2D) halide
  perovskites despite their unusually high exciton
  binding-energy.
  In this work we study the thermodynamic quasi-equilibrium of the
  relative population of
  photoexcited free quasi-electron/quasi-hole pairs,
  neutral-excitons and
  excitonic trions, in 2D materials that support such excitonic complexes with large binding energy.
  We derive and solve the general Saha equations which describe the detailed
  balance of such a system of photo-excited electronic degrees of freedom
  forming a multi-component fluid of excitations in thermodynamic
  quasi-equilibrium.
  The solution to these equations, for the special case of
  2D perovskites where the reported exciton and
  excitonic trion binding-energies are of the order of 0.3-0.4 eV for the former and 30-40 meV for the latter, reveals that
    while the charge-neutral excitonic population
    dominates all other excitations, at room
    temperature and below,  the excitonic trion
    component can be the dominant population among charge carriers. We also
    argue that trionic hopping can take
    place via a tunneling mechanism which is speculated
    to play a role in a novel charge-transport mechanism.
  \end{abstract}
\maketitle
\section{Introduction}
Understanding the character and behavior of the charge and energy flow in  semiconductor
quantum wells can lead to efficient optoelectronic materials.
In particular, the nearly ideal two-dimensional
(2D) Ruddlesden-Popper perovskites are quantum wells which are
self-assembled using wet-chemistry
synthesis\cite{Stoumpos2016,Mitzi1996}. Their  
band-gap is  tunable, by varying the perovskite-layer thickness,
and, thus, it can be used to modulate the effective electron-hole confinement.
One of their unique features is the large exciton binding energy\cite{Wu2015}
which, depending on the type of superlattice structure, can reach
values of several hundreds of meV. Analyzing the origin
and the consequences of this feature should play a significant role
towards understanding of the nature of charge carriers and transport.

The reduced screening in such 2D materials enhances the Coulomb interaction and leads to tightly bound exciton and other excitonic complexes.
For example, it has been
experimentally shown\cite{https://doi.org/10.1002/adma.202210221} that in ultra-thin layers of phenylethylammonium-lead-iodide
((PEA)$_2$PbI$_4$), the excitonic binding energy is approximately 0.35 eV.
It is widely expected that
light absorption creates excitons as long-lived excited states.
As a result their optical absorption spectrum  is dominated
by an intense exciton peak below the band edge, with estimated
binding energies of hundreds of meV. 
Therefore, in order to describe the equilibrium and non-equilibrium optical properties of
2D halide perovskites, first, we need to understand the
role of strongly bound excitons.

However, while
such a large exciton binding energy should hinder charge separation,
there is signiﬁcant evidence that there is an abundance of free carriers
when the material is photo-excited.
There are several attempts to explain such apparent contradiction.
One proposed explanation claims that the 
intra-gap edge-states formed at the crystalline grain
boundaries\cite{doi:10.1126/science.aal4211,https://doi.org/10.1002/adma.202201666,Zhao2019,Zhang2019} play a key role in exciton dissociation. This proposed mechanism is  related to the fact
that solution-processed thin films are typically polycrystalline with
grain boundaries featuring a high density of 
dangling bonds and defect edge-states.
Formation of polaronic excitons as well as
polarons, where the carriers are strongly coupled to lattice deformations,
are phenomena which seem to be established through various
experimental probes\cite{SrimathKandada2020,Yin2017,https://doi.org/10.1002/adom.202100295,Guzelturk2021,Zheng2016,Liu2020,https://doi.org/10.1002/adma.202007057,Yin2017,Tao2022,Franchini2021,Simbula2023} and they can play a significant role in
resolving the appearance of charge carriers.

In the present paper we calculate the populations 
of the various components of the photo-excited quasi-particle/quasi-hole
excitations, and, in addition, the neutral-excitonic and excitonic-trion
population of the electronic
system. We would like to point out that
exciton relaxation dynamics has been recently studied by 
 Ziegler {\it et al.}\cite{https://doi.org/10.1002/adma.202210221} in (PEA)$_2$PbI$_4$ where they demonstrate the emergence of both negatively and positively
charged excitonic trions, with binding energies up to 46 meV, among
the highest measured  in 2D systems. They also demonstrate that trions
dominate
light emission and propagate with mobilities reaching 200 cm$^2$V$^{-1}$s$^{-1}$ at
elevated temperatures.
The goal of our present paper is to study the statistical
mechanics of the quasi-equilibrium of a multi-component composite fluid of
the above mentioned long-lived photo-excitations in dynamic equilibrium with each other and with the gas of incident photons at a common temperature.
We involve two energy scales in the problem, the large
binding energy of an exciton which corresponds to several thousands
of K and a second energy scale which is the trion
binding energy, which is an order of magnitude smaller and corresponds
to hundreds of K. We derive and solve the Saha equations for
the composite fluid of these electronic excitations. We show that,
under broad-range of conditions, the trionic population
at room temperature and below  dominates the population of free quasi-electron
and quasi-hole carriers.

We also argue that these trionic charged excitations can move
through the lattice under the influence of an external electric field via a 
novel quantum-mechanical tunneling mechanism of trions mediated by
adjacent neutral-excitons. This mechanism relies on the
probability for a trion to be found next to a neutral exciton; however this
probability can become sizable at room temperature and below where
the population of neutral excitons is maximum because of their
large binding energy.

The paper is organized as follows. In the following section (Sec.~\ref{quasiparticles}) we discuss the nature of 
the photo-generated excitations, i.e.,  quasi-particle/quasi-hole
excitations and composite  quasiparticles
expected to be photo-created in 2D halide perovskites on general
quantum many-body theoretical grounds. In Sec.~\ref{simple-Saha}
we present and solve the Saha equations for the more familiar case
of neutral-excitons in equilibrium with photo-excited electron/hole pairs (which may be influenced by strong polaronic effects).
In Sec.~\ref{complex-Saha}
we derive  and solve the Saha equations for the more general case of a composite
fluid of neutral-excitons and positively and negatively charged excitonic trions in equilibrium with photo-excited quasi-electron/hole pairs.
In Sec.~\ref{tunneling} we discuss our speculation of
a trion-transport mechanism mediated by neutral excitons.
Lastly, in Sec.~\ref{conclusions} we present our conclusions.

\section{Nature of quasiparticles}
\label{quasiparticles}

There have been various attempts to demonstrate polaron-formation in the
2D halide perovskites \cite{SrimathKandada2020,Yin2017,https://doi.org/10.1002/adom.202100295,Guzelturk2021,Zheng2016,Liu2020,https://doi.org/10.1002/adma.202007057,Yin2017,Tao2022,https://doi.org/10.1002/adfm.202300363,Thouin2019,Franchini2021}.  For example, using high-resolution resonant impulsive stimulated Raman spectroscopy\cite{Thouin2019}, a vibrational wavepacket dynamics was identified that evolves along different configurational coordinates for distinct excitons and photo-carriers.
This observation\cite{Thouin2019} was interpreted as signature of
 the polaronic character
of excitons in two-dimensional lead halide
perovskites,
as different excitons induce specific lattice reorganizations.
Furthermore,  in Ref.~\onlinecite{SrimathKandada2020} it has been convincingly
argued that excitons, are in fact, exciton polarons, i.e., polaronic
effects play a significant role in the formation of 
bound states of electrons and holes dressed with their cloud of lattice distortions.  These effects are discussed in this section.

In addition, exciton relaxation dynamics has been recently studied\cite{https://doi.org/10.1002/adma.202210221} in (PEA)$_2$PbI$_4$ where the emergence of both negatively and positively
 charged excitonic trions, with binding energies up to 46 meV was discovered.
 Furthermore, it was found that excitonic trions
dominate
light emission and their mobilities reach 200 cm$^2$V$^{-1}$s$^{-1}$.

In this section, we first wish to  clarify various concepts regarding the nature of
elementary excitations
in these 2D insulators in a manner consistent with
quantum many-body theory.

\begin{figure}[!htb]
  \includegraphics[width=1.0\linewidth]{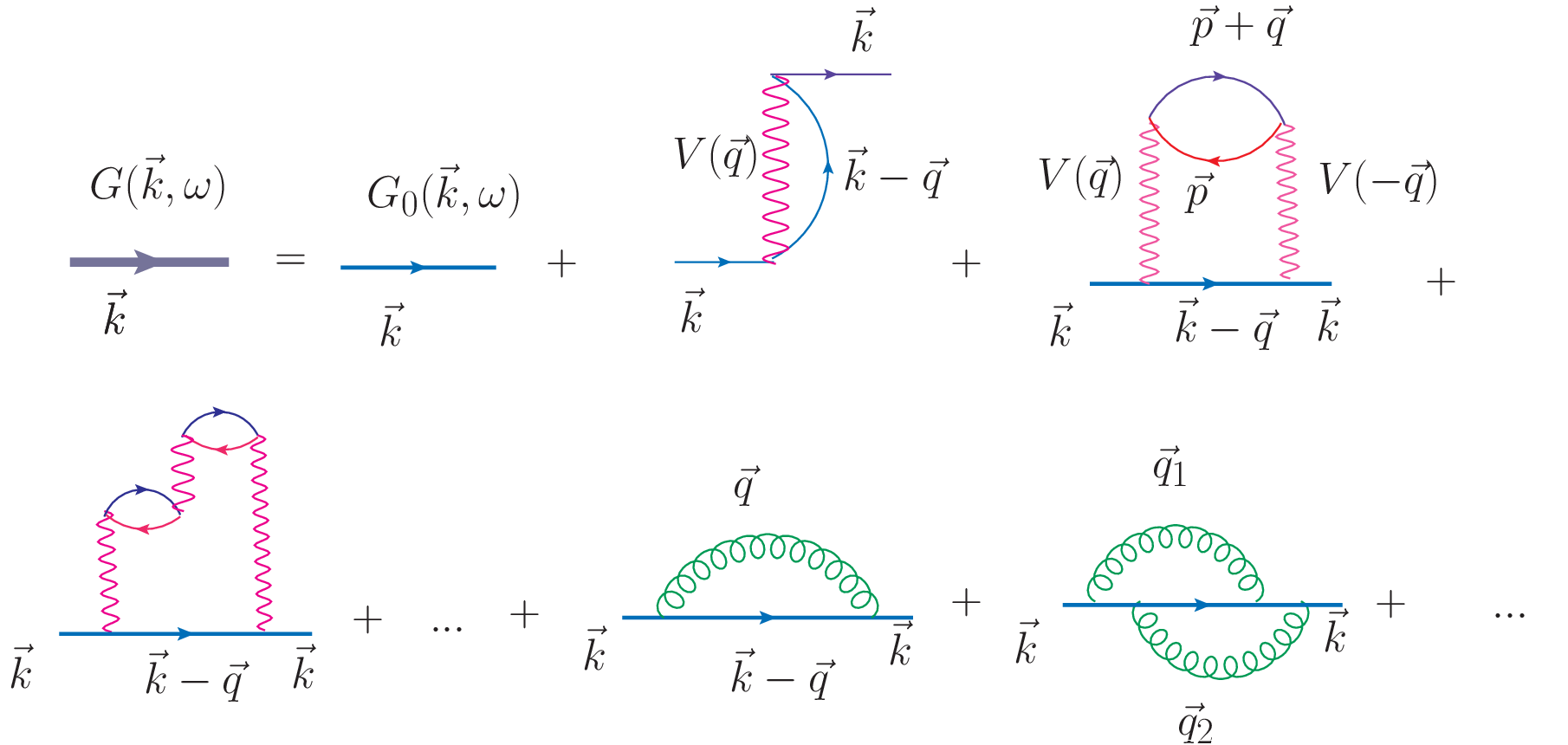}
  \caption{The interacting Green's function of the quasi-particles (and quasi-holes)
    which take into account the interaction with the other electrons
    of the Fermi sea and with the lattice. Its poles redefine
  the notion of the quasi-particle and quasi-hole to those which
  carry with them a cloud of virtual particle/hole excitations and
  a quantum gas of lattice virtual excitations (represented by the green
  wiggly-lines, i.e., polaronic effects).}
  \label{quasi-electron}
\end{figure}

When we think about bands in
materials, which are characteristic of the elementary excitations of the
many-body system, we begin from the non-interacting electron
represented by the bare propagator in Fig.~\ref{quasi-electron}.
This describes a single electron moving 
in some form of periodic pseudopotential due to the presence of the
 ions and
 some additional average (exchange-correlation) effective potential due to
 the spatially-dependent density-field generated by
 the collective presence of all the electrons.
 Such a picture can be best conceptualized within the Hohenberg-Kohn-Sham (HKS) density-functional-theory framework.
Even though this is conceptualized as
a free propagator, it includes the effects of the interactions with other electrons in  a way that gives meaning to the single-particle picture.
The HKS scheme is a purely ground-state theory.  In order to study excitations above such a ground-state
we need to include the effects of correlations. These effects
can be included perturbatively using
the residual screened Coulomb electron-electron interaction.
This interaction takes into account the effects of virtual particle/hole
excitations due to the Coulomb interaction.
In addition, allowing the ions to oscillate
around their equilibrium positions, i.e., by including the lattice
vibrations, leads to the quantization of their normal modes (phonons).
The effects of the coupling of the electrons to these lattice
excitations can be taken into account
either perturbatively, as shown in the second line of Fig.~\ref{quasi-electron}
or non-perturbatively in the strong coupling limit, using some deformation potential approach.
The series illustrated in Fig.~\ref{quasi-electron}
is written by means of the Dyson's equation for each band $n$ as follows:
\begin{eqnarray}
  G_n(\vec k,\omega) = {1 \over {\omega - e_n(\vec k) - \Sigma_n(\vec k, \omega)}}
\end{eqnarray}
where $e_n(\vec k)$ is the energy of the $n^{th}$ non-interacting band and
$\Sigma$ is the
so-called self-energy. The poles of this Green's function, assuming that
the imaginary part of $\Sigma$ becomes small as we approach the Fermi
surface, are the quasiparticles, which define the renormalized bands.
In the simplest picture, the ground-state corresponds
to filling these quasiparticle states with electrons up to the renormalized
Fermi energy. Beginning from this new ground-state one can carry out a
perturbative scheme with these new quasi-particles defined on
top of this Fermi sea and this scheme is repeated until
self-consistency is achieved\cite{PhysRevB.91.115105,PhysRevB.90.165142,PhysRevLett.96.226402,PhysRevLett.99.246403}. If this approach is met with no
singularities and anomalies, we can say that the Fermi-liquid picture
is valid and the band-picture has a meaning.

Polarons are a fancy way of characterizing the renormalization
of the quasi-electrons and quasi-holes which are also dressed by the virtually
excited phonons as shown by the green wiggly-line of Fig.~\ref{quasi-electron}.
Similarly, the electrons are dressed by the particle-hole excitations
shown by those diagrams which contain the magenta wiggly-interaction-line of Fig.~\ref{quasi-electron}.

After taking into account such  renormalization
effects, it leads to lowering of the energy of single-particle states near the Fermi level. Namely,
these effects  define the nature of the free-quasiparticle states out of which the bands and the ground-state are built.

These are the single quasi-particle and single quasi-hole states of this renormalized system of quasiparticles.
Namely, elementary excitations which are created
when a single electron or hole is added to the system.
However, the photo-excitations created by the incident light are based on 
the simultaneous creation of quasi-electron and quasi-hole excitations.
Initially, the incident light perturbs the
system and transfers energy to it. After a relatively
short-time
of the order of 1-100 fs, the excited state of the electronic system
can be described in terms of the following population 
of excitations. These excitation can be conceived as either free
quasi-electron/quasi-hole
pairs (i.e., electrons or holes dressed with their ground-state polarization clouds and the cloud of lattice distortions or lattice vibrations as in Fig.~\ref{quasi-electron})
or bound states of these renormalized quasi-electron/quasi-hole states.

    \begin{figure}[htp]
       \begin{center}
         \subfigure[]{
            \includegraphics[scale=0.5]{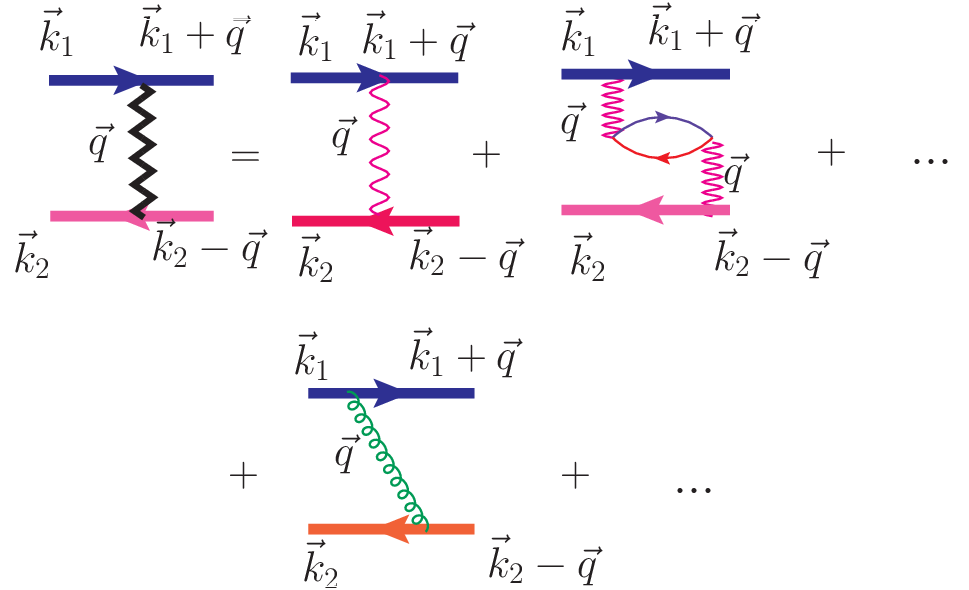} 
         }\\
         \subfigure[]{
               \includegraphics[scale=0.5]{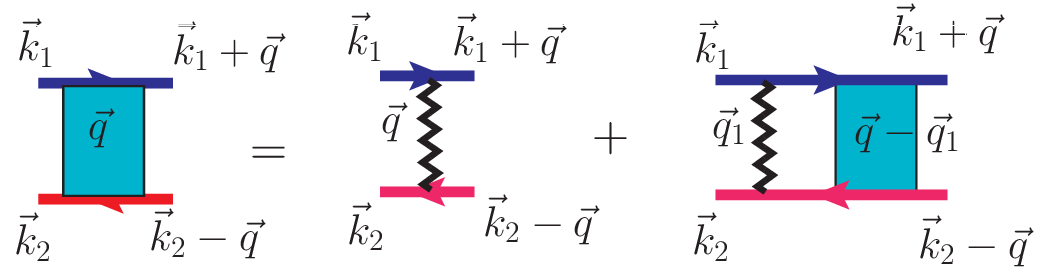} 
         }
       \end{center}
    \caption{(a) Renormalized quasiparticle interaction which takes the
    particle-hole polarization effects of the Fermi Sea within the RPA (terms
    containing the particle/hole bubble-like diagram) and
    the phonon-mediated interaction (green wiggly-line).
    (b) The Bethe-Salpeter equation for bound states between
    quasi-electrons and quasi-holes using the effective interaction defined
    in the (Top) part of this figure. The cyan box is the full two-body vertex in the  particle-hole channel.}
  \label{renormalizedinteraction}
\end{figure}

Once a quasi-electron (or polaron) is promoted from an occupied
renormalized  band to an unoccupied one,  the Coulomb interaction acts
between the created quasi-electron and the created quasi-hole.
This interaction is also renormalized by
including the effects of screening, namely, terms involving
the bubble-like diagram (Fig.~\ref{renormalizedinteraction}) where
quasi-particle/quasi-hole  pairs are virtually created. These effects
renormalize the
bare Coulomb interaction by means of the dielectric matrix within the
random-phase-approximation (RPA) and some
possible vertex corrections which go beyond RPA. In the case of the 2D
halide perovskites, this interaction is also significantly affected
by polaronic effects, which lead to a contribution 
to the residual interaction between quasi-particles (i.e., polarons)
which can be also attractive due to the effect of lattice
deformation. The leading perturbative term is shown in the top part of Fig.~\ref{renormalizedinteraction} as phonon-exchange interaction
between quasiparticles (green wiggly-line).
Once this interaction is constructed, we need to solve
the Bethe-Salpeter equation (BSE), i.e.,  the problem of a quasi-electron/hole  pair on top  of the interacting ground-state (in the
particle-hole channel)  as illustrated in
the bottom part of Fig.~\ref{renormalizedinteraction}.
We wish to emphasize that the quasi-particles entering the BSE can
be strongly renormalized polarons.
The solution to the BSE can lead to bound-states between a polaronic
quasi-electron and a polaronic quasi-hole, which are excitons.

Next, we consider the effects of the coupling to
phonons in an attempt to give a qualitative explanation of the results
reported in Ref.~\onlinecite{Thouin2019}, which were obtained by means of
high-resolution resonant impulsive stimulated Raman spectroscopy.
Let us just  consider the effects of the diagram involving the green wiggly-line
in Fig.~\ref{renormalizedinteraction} to the effective electron-hole
interaction, i.e., 
\begin{eqnarray}
  g^2_{\lambda}(\vec k,\vec q) {{\omega^2_{\lambda}(\vec q)} \over
      {\omega^2 - (\omega_{\lambda}(\vec q) - i \eta)^2}}
\end{eqnarray}
where $g_{\lambda}(\vec k,\vec q)$ is the electron-phonon coupling between
a phonon
mode $\lambda$ of frequency $\omega_{\lambda}(\vec q)$.
By examining this expression, we expect to have a stronger resonant
response in the excitonic spectrum at those phonon modes with
the largest electron-phonon coupling and for energy transfer
given by the phonon frequencies (we use units where $\hbar =1$).
While these experiments show that the electron
is dressed by the lattice vibrations, they do not prove that
a bound-state between an electron and lattice deformation occurs.
Moreover, the effect of the electron-phonon coupling may be taken into
account perturbatively; namely, this effect just leads to 
a renormalization of the standard notion of the quasi-electron
carrying with it the distortion of the background ionic-lattice.
In fact, because we are dealing with fast dynamics and the lattice
degrees of freedom respond with energy-scales of the order of a few
meV\cite{Thouin2019} (as expected on general grounds), 
 we should see the dressing effects
on the exciton after a time-scale of few ps following
the photo-excitation.
Therefore, nothing really unexpected was revealed by these experimental
findings suggesting that a dramatic non-perturbative effect
between electrons and lattice distortions happens in these materials.

Is it possible that the single-polaron state, carrying
with it a cloud of lattice deformations, as shown in Fig.~\ref{quasi-electron},
is a mobile bound-state to lattice distortions? Is it also possible that
its bound-state energy be lower than the energy of the excitonic state?
The answer is, in principle, yes to both questions. However, this would be an
extraordinary effect and, while there is strong evidence for
strong renormalization effects on the electronic spectrum due to
the virtual excitation of phonon modes\cite{Thouin2019}, there is no smoking gun indicating such a non-perturbative effect.

These excitonic bound states can have structure as shown 
in  Ref.~\onlinecite{SrimathKandada2020},
where the spectrum of
these exciton-polaron states is discussed.
As shown in Ref.~\onlinecite{SrimathKandada2020},
the center of the  excitonic energy is a few hundreds of
meV below the continuum of the free-carrier conduction-band.
In addition, there is a fine structure of various peaks separated by
a much finer energy difference  of the order of 30-40 meV.
The center of these excitations, i.e., a scale of few hundreds of meV
corresponds to
the exciton binding energy while the fine structure corresponds
to different types of excitons\cite{SrimathKandada2020}.

Now, once such excitonic bound states exist in the particle-hole channel,
which are
neutral excitations, we should ask if there are additional
bound-states, namely, charged excitations, which are bound states between
an exciton and a quasi-particle.
The reason for the possible existence of such excitations of bound states is
the fact that an electric-charge (that of the quasi-electron or
quasi-hole) and an electric-dipole (that of the exciton) interact.
In the formation of these {\it composite quasiparticles}, we expect the lattice to play a significant role, which means that polaronic effects are important
to obtain an accurate quantitative description.
Namely, they can be
excitonic-polaronic trions dressed with strong lattice deformations.
We discuss the nature of these bound-states, denoted $x^{\pm}$, in the following section (in Sec.~\ref{complex-Saha}).
In addition, we may have bi-exciton bound states which are also
strongly influenced by polaronic effects.

\section{Photons in equilibrium with excitons, electrons and holes}
\label{simple-Saha}
 \begin{figure}[htp]
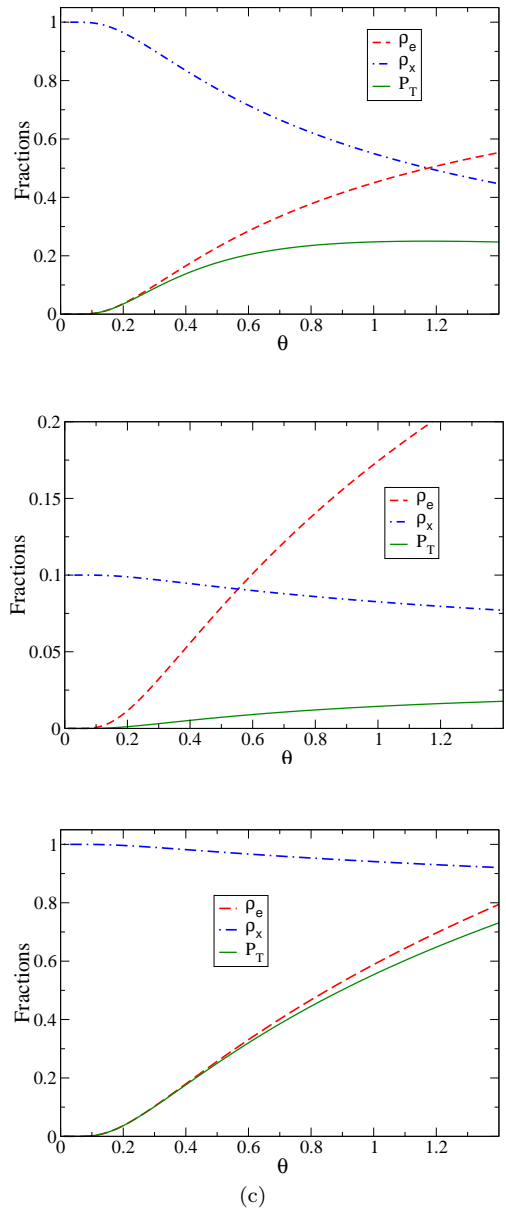

       \begin{center}
         \subfigure[]{
            \includegraphics[scale=0.27]{Fig3a.eps} 
         }\\
         \subfigure[]{
               \includegraphics[scale=0.27]{Fig3b.eps} 
         }\\
         \subfigure[]{
               \includegraphics[scale=0.27]{Fig3c.eps} 
         }
       \end{center}
\caption{Fractions of excitons, and electrons as a function of $\theta$
  for (a) $m^*_e=m^*_x=1$ and $\rho_p=1$, 
  (b) $m^*_e=m^*_x=10$ and $\rho_p=1$, 
  (c) $m^*_e=m^*_x=10$ and $\rho_p=10$} 
\label{excitons-and-carriers}
\end{figure}

When the incident photons are of energy greater than the 2D halide perovskite gap (of the order of 2 eV), they excite electrons from the valence band to the
conduction band (which is filled with quasi-electrons, i.e.,
electrons dressed with a cloud of lattice vibrations
and, to a degree of lesser importance, a polarization
cloud of virtual electrons and holes). After a time scale
of the order of 20-100 fs, form a population of free quasi-electron ($e^-$) and quasi-hole pairs
($h^+$) (i.e., quasiparticles of strong polaronic character) in addition to excitons ($x$) and excitonic trions $(x^{\pm})$ of strong polaronic
character. For simplicity in the rest of the this paper, we will refer to
these quasiparticles of electron or hole character as simply electrons (using
the symbol $e^-$), and holes (using the symbol $h^+$.

As a first step, we will ignore the presence of polaronic-trions
and we will deal
with the familiar case of excitons and electron/hole distributions.
Namely, we assume that every incident photon creates
an electron/hole pair some of which bind (within a short time scale
of the order of 10-100 fs) to form neutral excitons.
After a thermalization time-scale much shorter than the
recombination time-scale\cite{deQuilettes2019}
we have  photo-excited excitons ($x$) in quasi-equilibrium with photo-excited 
 electrons ($e^-$) and holes ($h^{+}$), i.e.,
\begin{eqnarray}
  {{x}} \longleftrightarrow  e^- + h^+,
\end{eqnarray}
with relative area densities of excitons ($n_x$) and
electron/hole pairs ($n_e=n_h$) satisfying the following constraint
\begin{eqnarray}
  n_p &=& n_x  + n_e,
\end{eqnarray}
where $n_p$ is the incident photon density.
In two dimensions we can describe this process by means of the so-called
Saha equations\cite{doi:10.1080/14786441008636148,doi:10.1098/rspa.1921.0029}. Detailed balance leads to the following
\begin{eqnarray}
  {{r_e
      r_h} \over {r_x}} &=&  {{m_e k_B T} \over {\pi \hbar^2}} e^{-{{{E^0_b} \over {k_B T}}}},\\
  r_e &=& {{n_e} \over {m^*_e}}, \hskip 0.1 in    r_h = {{n_h} \over {m^*_h}}, \hskip 0.1 in
  r_x = {{n_x} \over { m^*_x}}.
\end{eqnarray}
where $n_x, n_e, n_h$ are the average density (number of particles per unit area) of excitons, electrons and holes present in equilibrium ($n_e=n_h$).
Here $m_e^*$, $m^*_h$ and $m^*_x$ are
the effective electron, hole and exciton masses, respectively, in units of the
bare electron mass  $m_e$. The binding energy of the neutral exciton ($x$) is
denoted above as $E^0_b$.

 We can rewrite these equations in units of the constant
\begin{eqnarray}
  \kappa_0 \equiv {{E^0_b m_e} \over {\pi \hbar^2}},
\end{eqnarray}
which has dimensions of particle density, as
\begin{eqnarray}
  {{\rho_e
      \rho_h} \over {\rho_x}} &=&  \theta e^{-{{{1} \over {\theta}}}},\\
  \rho_p &=& \rho_x m^*_x + \rho_e m^*_e,\\
  \theta &\equiv& {{k_B T} \over {E^0_b}}, \hskip 0.1 in 
  \rho_e = {{r_e} \over {\kappa_0}}, \hskip 0.1 in    \rho_h = {{r_h} \over {\kappa_0}},
  \rho_x = {{r_x} \over {\kappa_0}}, 
\end{eqnarray}

This set of equations leads to the following solution for
the temperature dependence of the exciton and free-carrier fractions:
\begin{eqnarray}
  \rho_e &=& -{{\sigma_0(\theta)m^*_e} \over {2 m^*_x}} + \sqrt{\Bigl ({{\sigma_0(\theta) m^*_e} \over {2 m^*_x}}\Bigr )^2 + \sigma_0(\theta) {{\rho_p}\over {m^*_x}}},\\
  \rho_x &=& \rho_e^2/\sigma_0(\theta), \hskip 0.2 in
  \sigma_0(\theta) \equiv \theta e^{-{1 \over {\theta}}}
\end{eqnarray}
We note that $\sigma_0$ has dimensions of particle density.
These solutions are illustrated in Fig.~\ref{excitons-and-carriers},
where we can see that the relative number of photo-excited excitons
saturates to 1 slowly as a function of the temperature $\theta$.
The relative carrier number decreases with decreasing temperature as shown
on the bottom subfigure of Fig.~\ref{excitons-and-carriers}.

Notice that the curve labeled $P_T$ representing  the product
$\rho_e\rho_x$ which is proportional
to the probability to form a trion, is as sizable as
the fraction of the charge carriers below $\theta \sim 1$.
In the following Subsection we include explicitly the
population of trions by finding the Saha equations that correspond
to such case.

\section{Photons in equilibrium with excitons, trions, electrons and holes}
\label{complex-Saha}

\begin{figure}[!htb]
\includegraphics[width=0.7\linewidth]{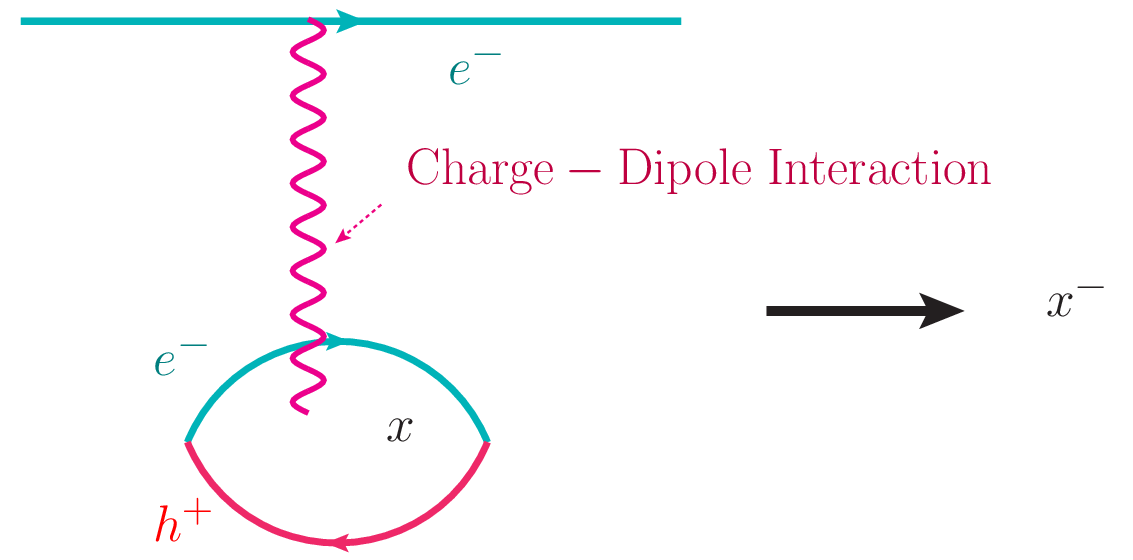} 
\caption{A photo-excited exciton ($x$) (illustrated schematically as
  a bubble formed by a quasi-electron and a quasi-hole propagator)
  with large binding energy in a
  semiconductor can attract a photo-excited
  quasi-electron (of polaron-like character as illustrated in Fig.~\ref{quasi-electron}) via the
  charge-dipole interaction (illustrated as magenta wiggly-line) to form a trion
  ($x^-$).
  The trion binding energy
  is expected to be weaker by an order of magnitude
compared to the exciton binding energy.}
\label{trion-formation}
\end{figure}

Fig.~\ref{trion-formation} illustrates that the presence of a residual attractive
charge-dipole interaction between an exciton, i.e., a bound electron-hole pair
and a photo-excited quasiparticle (with strong polaronic character) in the conduction band.
Under favorable conditions this interaction can lead to a bound state
which for simplicity we call it a negatively charged excitonic trion.
There is computational\cite{PhysRevLett.126.216402} and experimental\cite{https://doi.org/10.1002/adma.202210221,10.1063/1.5125628} evidence
that layered halide perovskites host such states and their binding energy is
typically an order of magnitude smaller than that of excitons, i.e.,
in the tens of meV range\cite{PhysRevLett.126.216402}.
When this state dissociates it dissociates to a polaron
and a neutral
exciton. It is expected to be quasi-stable for some time longer
than the time required to reach quasi-equilibrium with the rest of the
carriers and excitons.

We will also assume
that the recombination time is much longer\cite{deQuilettes2019,Brennan2018,BRENES2017} than the  time required
for the gas of these
excitations, i.e., quasi-electrons ($e^-$) and quasi-holes ($h^+$),
excitons ($x$) and excitonic trions
(which we generically denote as $x^-$ and $x^+$) to
form a transient quasi-equilibrium. When this quasi-equilibrium is reached,
let the  area densities of these excitations be, $n_e$, $n_h$, $n_x$, $n_{x^-}$ and $n_{x^{+}}$ respectively.

We will also assume that initially every incident photon creates one
electron/hole pair, a fraction of which rather quickly combine to form
excitons ($x$), and charged excitonic trions ($x^{\pm}$). When
quasi-equilibrium is reached via the following processes,
\begin{eqnarray}
  {{x}} \longleftrightarrow    e^-  + h^+, \label{bal1}\\
    {{x}^{-}} \longleftrightarrow  {{x}} + e^-,\label{bal2} \\
   {{x}^{+}} \longleftrightarrow {{x}} + h^+, \label{bal3}
\end{eqnarray}
the equilibrium densities are related to the photon-density
$n_p$ as follows:
\begin{eqnarray}
  n_p = n_x &+& {3 \over 2} ({n_{x^-} + n_{x^+}}) + {1 \over 2} ({n_e + n_h}),\label{densities}\\
  n_{x^-} + n_e &=& n_{x^+} + n_h,\label{neutrality}
\end{eqnarray}
where the last equation is the statement of charge neutrality.  Eq.~\ref{densities}  is obtained by assuming that every incident photon initially
creates one electron/hole pair and at equilibrium  $n_{x^-}$ of the electrons
combine with equal number of neutral excitons to form $n_{x^-}$ 
negatively charged excitonic trions while $n_{x^+}$ of these
initially photo-generated holes combine with an equal number of
neutral excitons to form $n_{x^+}$ of positively charged excitonic trions.
Eq.~\ref{densities}
is derived in Appendix~\ref{density-relations}.

The object on the left-hand-side (LHS) of each one of the above balance
equations (Eqs.~\ref{bal1},\ref{bal2},\ref{bal3})
is a bound-state of the two objects on the
right-hand-side (RHS) of the same equation, therefore, it has a lower energy.
The excitation from the object on the LHS
to that on the RHS and vice versa should satisfy
detailed balance.  Equating the corresponding product of the equilibrium canonical distribution with the transition probability and that of the
inverse process, at a common
temperature $T$, leads to the following set of coupled Saha-type equations:
\begin{eqnarray}
  {{r_e
      r_h} \over {r_x}} &=& \sigma_0  \hskip 0.2 in
  {{r_x r_e} \over {r_{x^{-}}}} = \sigma_{-} \hskip 0.2 in
  {{r_x r_h} \over {r_{x^{+}}}} = \sigma_{+} \\
    r_{\lambda} &=& {{n_{\lambda}} \over {m^*_{\lambda}}},\hskip 0.2 in \lambda=x,x^{-},x^{+},e,h, \\
    \sigma_{\tau} &\equiv & {{m_e  k_B T} \over {\pi \hbar^2}} e^{-{{|E^{\tau}_b|} \over {k_B T}}}, \hskip 0.2 in \tau = 0, \pm,
    \end{eqnarray}
where $m^*_{x^{\pm}}$ is  the effective mass (in units of the bare electron mass
$m_e$) of the trions, and $m^*_x,m^*_e,m^*_h$ are the effective masses of the excitons, electrons  and holes. The area densities $n_p, n_x, n_{x^-}, n_{x^+},  n_e, n_h$ have been defined earlier (after Eq.~\ref{densities},\ref{neutrality}).
Here, $E^{\pm}_b$ are the trion binding energies relative to the exciton binding
energy.

Using the parameters $\theta$ and $\kappa_0$ introduced earlier we can rewrite these equations as follows
\begin{eqnarray}
  {{\rho_e
      \rho_h} \over {\rho_x}} &=& \sigma_0(\theta),  \hskip 0.1 in
  {{\rho_x \rho_e} \over {\rho_{x^{-}}}} = \sigma_{-}(\theta), \hskip 0.1 in
  {{\rho_x \rho_h} \over {\rho_{x^{+}}}} = \sigma_{+}(\theta), \label{excitonic}\\
    \rho_{\lambda} &=& {{r_{\lambda}} \over {\kappa_0}}, \hskip 0.1 in
    \sigma_{\pm}(\theta) \equiv  \theta e^{-{{\alpha^{\pm}} \over {\theta}}}, \hskip 0.1 in  \alpha_{\pm} = {{E^{\pm}_b} \over {E^0_b}},
\end{eqnarray}
which should satisfy Eqs.~\ref{densities},\ref{neutrality}, i.e., 
\begin{eqnarray}
 \rho_p &=& \Bigl [{m_x^*}  + {3 \over {2}}
  (\xi_-+ \xi_+)\Bigr ]{{\rho_e\rho_h} \over {\sigma_0}} + {{m^*_e\rho_e+m^*_h\rho_h} \over 2},\hskip 0.2 in \label{eqn1}\\
 &&     (\xi_--\xi_+) \rho_e\rho_h =
          \sigma_0  ( m^*_h \rho_h- m^*_e \rho_e ), \label{eqn2} 
\end{eqnarray}
where
\begin{eqnarray}
  \xi_- = {{m^*_{x^-}\rho_e} \over {\sigma_-}}, \hskip 0.2 in
  \xi_+ = {{m^*_{x^+}\rho_h} \over {\sigma_+}}.
\end{eqnarray}
Eqs.~\ref{eqn1},\ref{eqn2}
can be solved in terms of $\rho_e$ and $\rho_h$, where
the reduced photon density $\rho_p$ and the effective masses are taken as
input parameters.
After obtaining these two densities, the excitonic and trionic densities
are found using Eqs.\ref{excitonic}.

 \begin{figure}[htp]
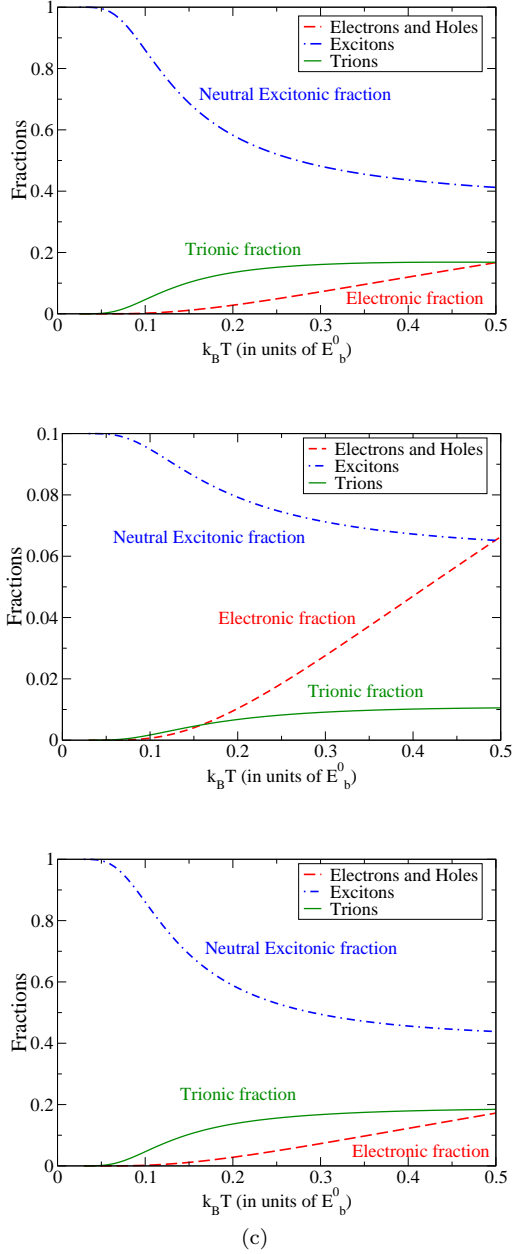

       \begin{center}
         \subfigure[]{
            \includegraphics[scale=0.27]{Fig5a.eps} 
         }\\
         \subfigure[]{
               \includegraphics[scale=0.27]{Fig5b.eps} 
         }\\
         \subfigure[]{
               \includegraphics[scale=0.27]{Fig5c.eps} 
         }
       \end{center}
\caption{Temperature dependence of the fractions of the various components
  for the case of $m^*_e=m^*_h=m^*_x=m^*_{x^-}=m^*_{x^+}$ and $\rho_p=1$ (a).
$m^*_e=m^*_h=1$, $m^*_x=m^*_{x^-}=m^*_{x^+}=10$ and $\rho_p=1$ (b).
  $m^*_e=m^*_h=1$, $m^*_x=m^*_{x^-}=m^*_{x^+}=10 m^*_e$ and $\rho_p=10$
(c).}
\label{carrier-densities}
\end{figure}

In order to see the qualitative
behavior of the solution to the above equations, we take the case where
$E^+_{b}=E^{-}_b$ and $m^*_{x^+}=m^*_{x^+}=m_e=m^*_h=m^*_e$, which imply that $n_{x^-}=n_{x^+}$.
In Fig.~\ref{carrier-densities} (top) we plot the fractions of the various
carriers as a function of $T$ in units of $E^0_b$ (i.e., as a function of
$\theta$)
for the case where $\alpha_{\pm} =0.1$, using photon density $\rho_p=1$ (in units of $\kappa_0$).
Notice that the trionic density remains constant as the temperature
is lowered at and below some temperature-scale of the order of $E^{\pm}_b$ (which is an order of magnitude smaller than the excitonic binding energy $E^0_b$)
the trionic density begins to drop.
In Fig.~\ref{carrier-densities} (middle) we plot the fractions of the various
carriers as a function of $\theta$
for $\alpha_{\pm} =0.1$, using $m^*_{x^+}=m^*_{x^+}=m^*_x=10$ and $m_e=m^*_h=m^*_e=1$
and photon density
$\rho_p=1$. Note the significant reduction to both the excitonic and trionic
densities. However, if we increase the photon density to $\rho_p=10$,
the density of excitons and trions (Fig.~\ref{carrier-densities} (bottom)) increases to their initial values (i.e., Fig.~\ref{carrier-densities} (top)).

To understand these values in physical units, $\rho_p$ is in units of
$\kappa_0 \simeq 7 \times 10^5 \mu m^{-2}$ and when the incident  light
is green and is under solar illumination fluences, which are orders of magnitude
lower excitation fluences than in most spectroscopic measurements,
the photon density per unit time is
2.5 $\times 10^{9}$ photons/($\mu$m$^2$sec). The photocarrier decay times in well-passivated perovskites have been reported to be as long as 10$\mu s$,   as measured by both photo-luminescence and microwave
conductivity\cite{Brennan2018,BRENES2017}.
This implies that in units of $\kappa_0$, the average total carrier density
is $\rho_p \sim 0.03$. Therefore, the values of $\rho_p=1,10$ used in
Fig.~\ref{carrier-densities} should be in the range of typical excitation fluences used in the reported spectroscopic measurements.

\section{Trion tunneling}  
\label{tunneling}

\begin{figure}[!htb]
  \includegraphics[width=0.9\linewidth]{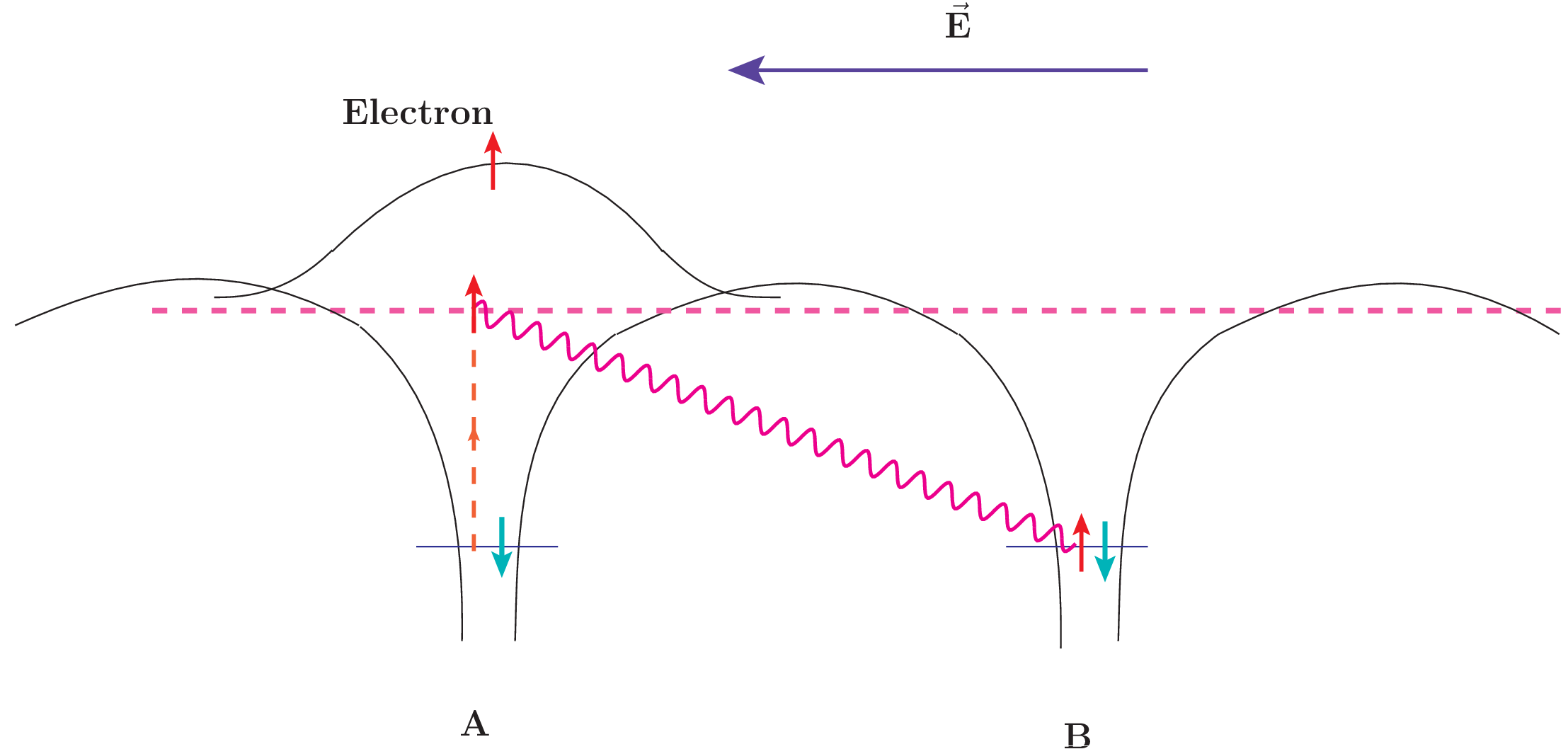} \\
  \vskip 0.2 in
\includegraphics[width=0.9\linewidth]{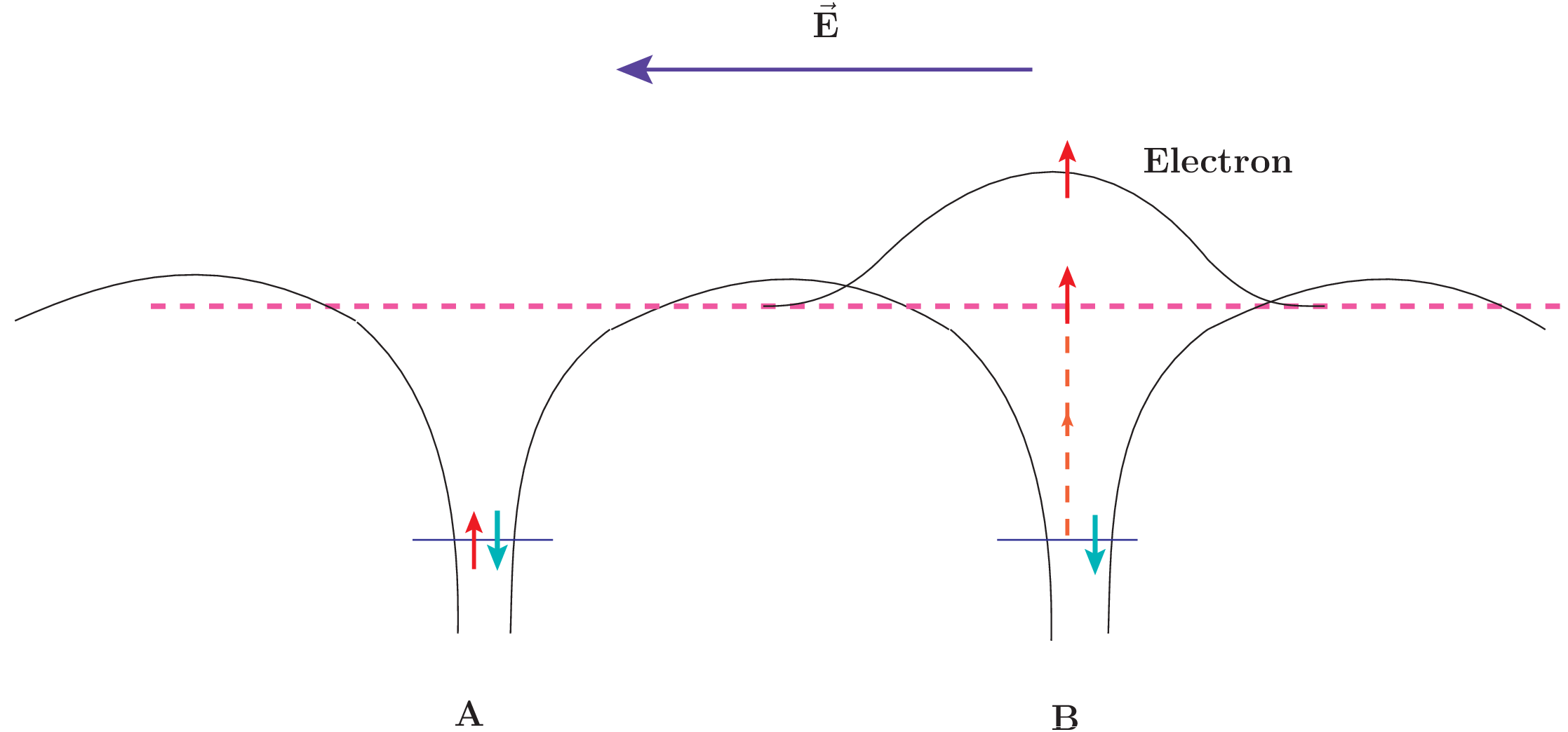}\\
\caption{ (Top) 
  A third electron (schematically shown
  by the wavefunction envelope with an up-spin electron occupying it in site A) bound to an exciton (schematically illustrated in
  real space by the red arrow indicating
  that the up-electron was promoted to an excited Wannier orbital)
  forming a trion.  For ordinary trion-tunneling to occur to a nearest neighboring site,  simultaneous energy transfer through the Coulomb interaction (magenta wiggly line) and electron transfer to a nearest neighboring atomic site
  is required. (Bottom) When these requirements are simultaneously met, the electron (and, thus, the trion) transfers to that nearest-neighbor site.} 
\label{trion-tunneling}
\end{figure}

Fig.~\ref{trion-tunneling} illustrates that
an electrically charged excitonic state (a trion) in the presence of an external electric
  field could tunnel from one atomic site to a nearest neighboring atomic
  site. This quantum-mechanical tunneling process is enabled by the
  energy offset between neighboring sites, which is caused by the presence
  of the electric field, ${\vec E}$, that lowers the energy of neighbor sites by an amount 
  $-e {\vec E} \cdot  \delta {\vec  R}$ (where $\delta {\vec  R}$ is the difference between the Bravais
  lattice vectors of two neighboring sites). This particular trion-tunneling mechanism requires a combined process to occur.  First, the energy of the excited exciton, i.e., the
  bound electron/hole pair at the original site (to which the additional charge is bound), is transferred to the
  nearest-neighbor site by means of the
  electron-electron interaction (magenta wiggly-line in Fig.~\ref{trion-tunneling}), thereby creating a neutral exciton there. 
  Simultaneously, the additional third electron bound to the trion $x^-$ follows
  under the influence of the external electric field. Because this
  is a coherent combination of two processes, it is proportional to
  the product of the amplitude for the energy transfer and the tunneling amplitude for the electron to follow.

\begin{figure}[!htb]
  \includegraphics[width=0.9\linewidth]{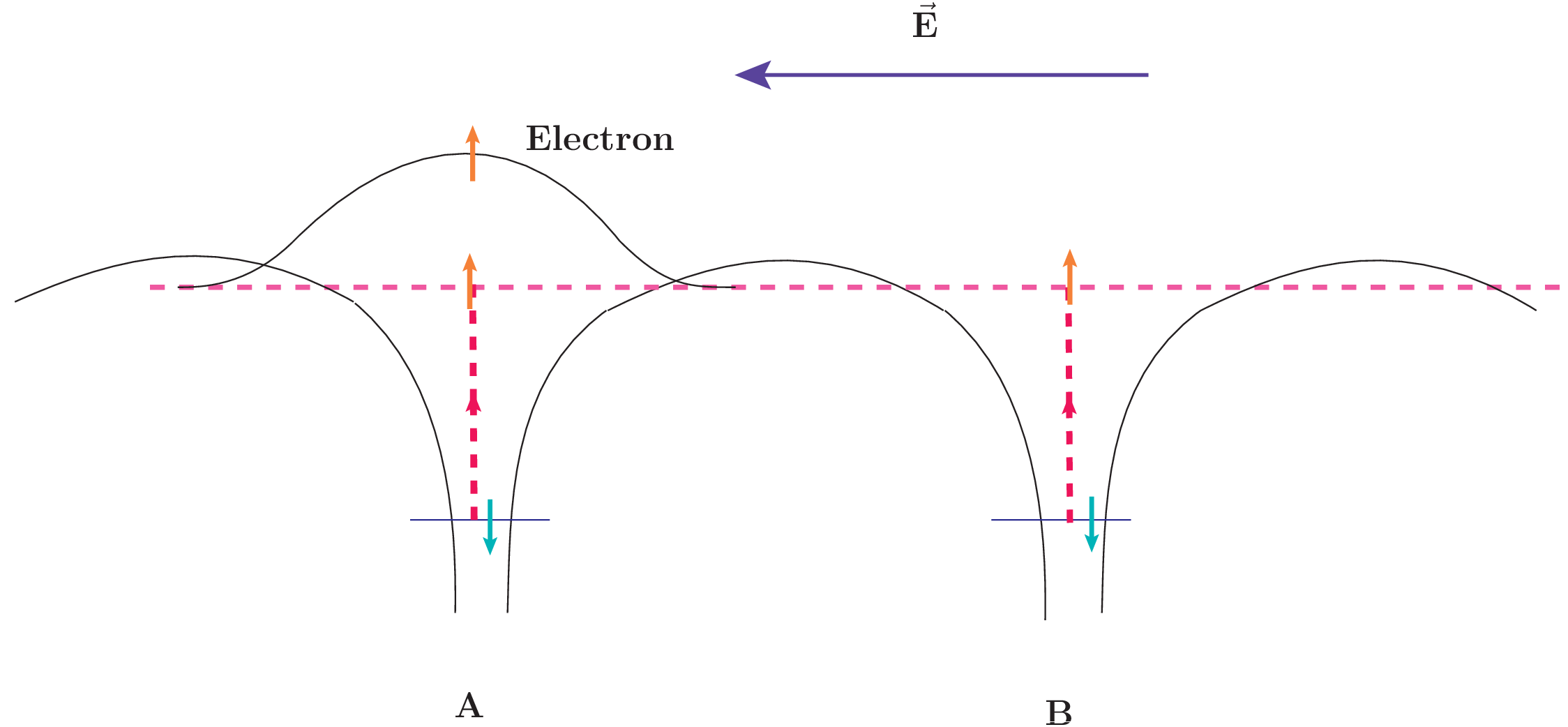} \\
  \vskip 0.2 in
\includegraphics[width=0.9\linewidth]{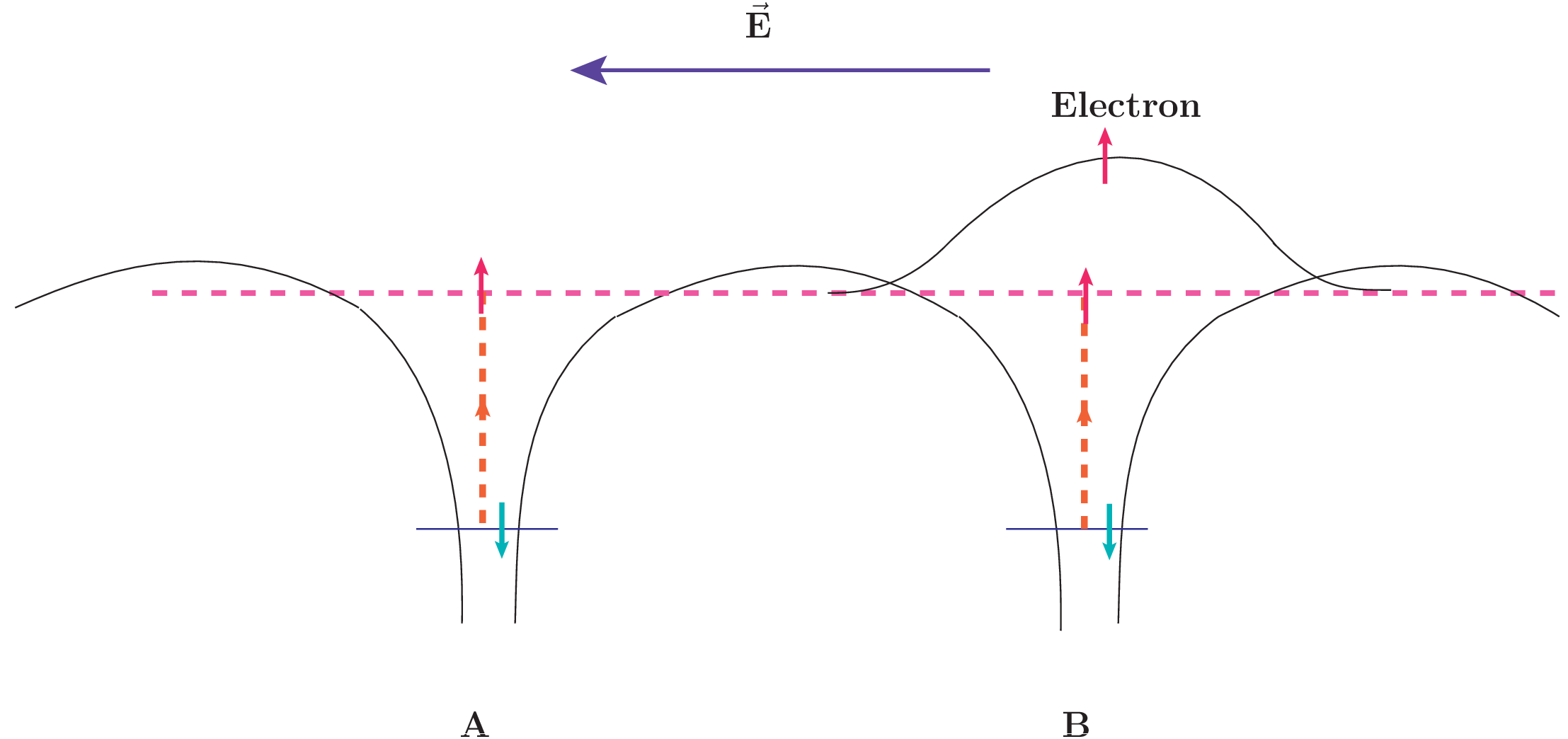}
\caption{(Top) Since the neutral-exciton density can be very high at room temperature and below (Fig.~\ref{carrier-densities}), there is significant probability
  that the third electron (schematically shown
  by the wavefunction envelope with up-spin electron occupying it in site A) bound to an exciton (schematically illustrated in
  real space by the red arrow indicating
  that the up-electron was promoted to an excited Wannier orbital)
  forming a trion  to find a neutral
  exciton next to it in site B. (Bottom) This will allow trion tunneling to such neighboring site
  by only charge (electron) hopping to site B producing the
  state shown in the bottom subfigure.}
\label{trion-tunnelingb}
\end{figure}

Notice that simultaneous energy and charge transfer is necessary to obtain
hopping of a trion from one site to its nearest neighbor
(Fig.~\ref{trion-tunneling}). However,  there is the possibility that the
neighboring site to a trion be occupied by an exciton (Fig.~\ref{trion-tunnelingb}). As we have
shown the neutral exciton density (depending on the
illumination fluences and recombination rate) can be high at or below room
temperature
(for room temperature $k_BT << E^0_b$, because the exciton binding energy $E^0_b$ is 0.35 eV), therefore, this becomes probable. In this case, the tunneling
process in the presence of an exciton next to a trion, as illustrated
in Fig.~\ref{trion-tunnelingb}, requires only charge-transfer without energy-transfer, which
can become more likely than the case of Fig.~\ref{trion-tunneling}.
 A very similar process can take place which leads to the tunneling of
  $x^+$ trions. By means of such 
  processes trions of  positive charge $x^{+}$, i.e., bound-states of an exciton with a photo-excited hole, under the influence of an external electric field
  will tunnel in the opposite direction and  be collected by the opposite polarity lead.

Therefore, there is a quantum mechanical mechanism for trion
transfer, which does not fall into the semi-classical mechanism of electrical
transport and could play a role in the
photo-conductivity mechanism in these materials.

\vskip 0.5 in

\section{Discussion and Conclusions}
\label{conclusions}

In Sec.~\ref{quasiparticles} we have discussed the nature of the
the elementary quasiparticles as well as that of composite quasiparticles which are expected to be photo-generated in 2D halide perovskites.

The relative populations of excitonic complexes which contain neutral excitons
($x$) as well as negatively ($x^-$) as well as positively charge ($x^+$)
excitonic trions in quasi-equilibrium with the photo-excited electron-hole
pairs as a function of temperature $T$ was studied in the limit
where the binding energy for $x$ excitons is a significant fraction of one eV
and that for $x^{\pm}$ is of the order of tens of meV.
The Saha equations are derived for this multi-component system of
assumed long-lived photo-excitations in equilibrium with an incident photon
gas of fixed area-density $n_p$. This is a well-defined problem of
Statistical Physics of very little, if any, ambiguity.
In addition to the above mentioned  energy scales, the only other
parameters entering the problem are the effective masses of the components
of the system and $n_p$.

The solution to this problem reveals that as a function of temperature
at room temperature and below for a wide-range of values of the above mentioned effective masses, for sufficiently high values of $n_p$, the trionic population
is the dominant charge excitation of this system.

These findings, in conjunction with recent experimental evidence
for their presence, the behavior and mobility  of such excitonic-trions
in 2D hybrid perovskites\cite{https://doi.org/10.1002/adma.202210221,10.1063/1.5125628} can explain the observed seemingly contradictory behavior of
these materials, namely, that despite their large excitonic binding
energy, there are plenty of charge carriers.

In this paper, it is also speculated in Sec.~\ref{tunneling} that, when there is an applied external electric, there are
two different mechanisms of trion hopping to a nearest neighboring site.
One is a spontaneous process involving both energy and charge transfer and
a second is a hopping process stimulated by a high neutral-exciton population, as expected to be the case at or below room temperature
(when the incident-light fluences are high enough), in which case
trion tunneling is mediated by a nearest neighboring neutral-exciton.
This speculation can help explain the
observed sizable trion mobility\cite{https://doi.org/10.1002/adma.202210221,10.1063/1.5125628}, despite their complex structure and leads to
a novel transport mechanism in these materials.
Excitonic trions were recently  invoked to explain quantum beating phenomena in bulk perovskites\cite{Huynh2022} and in applications of time-resolved spectroscopic techniques to quantum dots\cite{Makarov2016} and
colloidal nanocrystals\cite{Raino2016} of cesium–lead-halide and as
well as in attempts to analyze the excitation fluences dependence of transient absorption signals
in CsPbBr$_3$\cite{Nakahara2018}.

At a time-scale where recombination occurs, it is reasonable to expect
that these
excitonic complexes, both the neutral and the charged excitons,
will decay into photons and give rise to photo-luminescence.

The role of biexcitons\cite{PhysRevMaterials.2.034001,Su2017,https://doi.org/10.1002/adom.201800616} can  also be taken into consideration by extending the present approach. However,  we tried to minimize the number of parameters involved in the present study for clarity.  We do expect biexcitons to contribute with
a somewhat sizable population at and below room temperature by depleting
the exciton population by a very small fraction.
However, we have no reason to expect that their inclusion would
alter the qualitative results of our study regarding the relative populations
of the charged excitations, i.e., trions and electron/hole pairs.

Finally, the scenario and an extension of the model of the role of trions  discussed here, and in particular the Saha equations, which describe  the quasi-equilibrium of a fluid
of photo-excitations (i.e., unbound quasi-electron/quasi-hole pairs,
excitons and trions) 
may be applicable to other materials. In particular, we expect this work
to find application in transition metal
dichalcogenides, where observations that these materials host exciton and trions
with large binding energy have already been reported\cite{Li2018,PhysRevB.97.195408,Zhao2024,Chen:23}.

\section{Acknowledgments}
The author is thankful to Hanwei Gao and Peng Xiong for useful discussions
and for sharing their photo-conductivity measurements prior to
publication. The author is also grateful to Peter Rosenberg for
proofreading the manuscript.
This work was supported by the U.S. National Science Foundation under Grant No. NSF-EPM-2110814.
\appendix
\section{Derivation of relationship between excitation densities}
\label{density-relations}

In order to balance the number of absorbed photons per unit volume, $n_p$,
we will breakdown the thought process as a three level counting process.
First, let us say that every absorbed photon creates an unbound electron/hole
(e/h) pair, i.e.,
\begin{eqnarray}
  n_p = n^0_{e/h}.
  \end{eqnarray}
However the number of unbound e/h pairs, i.e.,  $n^0_{e/h}$, is the same as the number of unbound electrons or the number of unbound holes,
i.e., $n^0_{e/h}=n^0_e=n^0_h=(n^0_e+n^0_h)/2$, and in order to
have a particle-hole symmetric equation, we can write:
\begin{eqnarray}
  n_p = {{n^0_{e}+n^0_{h}} \over 2}\label{eq0}.
\end{eqnarray}
Next, we allow equilibrium between unbound e/h pairs and neutral
excitons $x$, i.e., the process
\begin{eqnarray}
  {{x}} \longleftrightarrow    e^-  + h^+, 
\end{eqnarray}
to take place. At the end of this process, a number $\delta n^0_e$ of electrons
and an equal number of holes $\delta n^0_h$ ($\delta n^0_e=\delta n^0_h$)
will bind to form an equal number $n_x=\delta n^0_e=\delta n^0_h$ of neutral excitons. The right-hand-side of the above Eq.~\ref{eq0} will be transformed
as
\begin{eqnarray}
  n_p = n^{(1)}_x + {{n^{(1)}_{e}+n^{(1)}_{h}} \over 2}\label{eq1}.
\end{eqnarray}
Namely, straightforward
substitution of $n^{(1)}_e=n^0_e-\delta n^{(0)}_e$, $n^{(1)}_h=n^0_h-\delta n^{(0)}_h$ and
$n^{(1)}_x=\delta n^{(0)}_e = n^{(1)}_e=n^0_e-\delta n^{(0)}_h$ in the above equation yields
the originally assumed Eq.~\ref{eq0}.

In the third stage, we continue to keep the number of photons unchanged and by allowing the right-hand-side
of Eq.~\ref{eq1}
to change by using $x$'s and $e^-$'s and $h^+$'s to create $x^-$'s and $x^+$'s, namely the following equilibrium processes
\begin{eqnarray}
    {{x}^{-}} \longleftrightarrow  {{x}} + e^-, \\
   {{x}^{+}} \longleftrightarrow {{x}} + h^+. 
\end{eqnarray}
Let us say  $\delta n^{(1)}_x$ of the $x$'s are transformed to $x^-$'s to produce a number $n_{x-}$ of the latter, so that
\begin{eqnarray}
  \delta n^{(1)}_x=n_{x-}.\label{eq2}
  \end{eqnarray}
However,  they need to combine with $\delta n^{(1)}_e$ electrons from the $n^{(1)}_e$ electrons to do it. Therefore, they will reduce the number of
unbound electrons by:
\begin{eqnarray}
  \delta n^{(1)}_e = \delta n^{(1)}_{x} \label{eq3}
\end{eqnarray}
and so the remaining electron-density $n_e$ will be given by
\begin{eqnarray}
n_e = n^{(1)}_e - \delta n^{(1)}_e = n^{(1)}_e - \delta n^{(1)}_x.\label{eq3b}
\end{eqnarray}

Similarly, let us say another part $\delta n^{(2)}_x$ of the $x$'s is
transformed to $x^+$'s to produce a number of  $n_{x+}$ of the
latter, so that
\begin{eqnarray}
  \delta n^{(2)}_x=n_{x+}.\label{eq4}
  \end{eqnarray}
There neutral excitons will need to combine with
$\delta n^{(1)}_h$ holes from the $n^{(1)}_h$ holes to do it.
Therefore, they will reduce the number of
unbound holes by an amount of
\begin{eqnarray}
 \delta n^{(1)}_h = \delta n^{(2)}_x.\label{eq5}
\end{eqnarray}
Hence, the remaining hole-density $n_h$ will be given by
\begin{eqnarray}
n_h = n^{(1)}_h - \delta n^{(1)}_h = n^{(1)}_h - \delta n^{(2)}_x.\label{eq5b}
\end{eqnarray}
The right-hand-side of Eq.~\ref{eq1} is now written as
\begin{eqnarray}
n_p &=& n_x + \delta n^{(1)}_x + \delta n^{(2)}_x + {{n_e + n_h} \over 2}\nonumber \\
  &+& {{\delta n^{(1)}_e + \delta n^{(1)}_h} \over 2}.
\end{eqnarray}
Namely, we substituted in Eq.~\ref{eq1} that $n^{(1)}_x=n_x + \delta n^{(1)}_x +
\delta n^{(2)}_x$,
$n^{(1)}_e = n_e + \delta n^{(1)}_e$ and 
$n^{(1)}_h = n_h + \delta n^{(1)}_h$.

Next, we substitute for $\delta n^{(1)}_e$ from Eq.~\ref{eq3}, and for
$\delta n^{(1)}_h$ from Eq.~\ref{eq5}
to obtain
\begin{eqnarray}
n_p &=& n_x + \delta n^{(1)}_x + \delta n^{(2)}_x + {{n_e + n_h} \over 2}\nonumber \\
  &+& 
   {{ \delta n^{(1)}_x + \delta n^{(2)}_x} \over 2},
\end{eqnarray}
which leads to 
\begin{eqnarray}
n_p = n_x  + {{n_e + n_h} \over 2}
  + {{3(\delta n^{(1)}_x + \delta n^{(2)}_x)} \over 2}.
\end{eqnarray}
We now use the fact that $\delta n^{(1)}_x= n_{x_-}$ (Eq.~\ref{eq2}) and $\delta n^{(2)}_x= n_{x_+}$ (Eq.~\ref{eq4}) to write
\begin{eqnarray}
n_p = n_x  + {{n_e + n_h} \over 2}
  + {{3(n_{x_-} + n_{x_+})} \over 2},
\end{eqnarray}
which is Eq.~\ref{densities} used in the main part of the paper.

Note that this final equation does not depend on the order of 
the various transformation processes. It simply expresses
the pseudo-particle number conservation in these transformations.


\begin{thebibliography}{42}%
\makeatletter
\providecommand \@ifxundefined [1]{%
 \@ifx{#1\undefined}
}%
\providecommand \@ifnum [1]{%
 \ifnum #1\expandafter \@firstoftwo
 \else \expandafter \@secondoftwo
 \fi
}%
\providecommand \@ifx [1]{%
 \ifx #1\expandafter \@firstoftwo
 \else \expandafter \@secondoftwo
 \fi
}%
\providecommand \natexlab [1]{#1}%
\providecommand \enquote  [1]{``#1''}%
\providecommand \bibnamefont  [1]{#1}%
\providecommand \bibfnamefont [1]{#1}%
\providecommand \citenamefont [1]{#1}%
\providecommand \href@noop [0]{\@secondoftwo}%
\providecommand \href [0]{\begingroup \@sanitize@url \@href}%
\providecommand \@href[1]{\@@startlink{#1}\@@href}%
\providecommand \@@href[1]{\endgroup#1\@@endlink}%
\providecommand \@sanitize@url [0]{\catcode `\\12\catcode `\$12\catcode
  `\&12\catcode `\#12\catcode `\^12\catcode `\_12\catcode `\%12\relax}%
\providecommand \@@startlink[1]{}%
\providecommand \@@endlink[0]{}%
\providecommand \url  [0]{\begingroup\@sanitize@url \@url }%
\providecommand \@url [1]{\endgroup\@href {#1}{\urlprefix }}%
\providecommand \urlprefix  [0]{URL }%
\providecommand \Eprint [0]{\href }%
\providecommand \doibase [0]{https://doi.org/}%
\providecommand \selectlanguage [0]{\@gobble}%
\providecommand \bibinfo  [0]{\@secondoftwo}%
\providecommand \bibfield  [0]{\@secondoftwo}%
\providecommand \translation [1]{[#1]}%
\providecommand \BibitemOpen [0]{}%
\providecommand \bibitemStop [0]{}%
\providecommand \bibitemNoStop [0]{.\EOS\space}%
\providecommand \EOS [0]{\spacefactor3000\relax}%
\providecommand \BibitemShut  [1]{\csname bibitem#1\endcsname}%
\let\auto@bib@innerbib\@empty
\bibitem [{\citenamefont {Stoumpos}\ \emph {et~al.}(2016)\citenamefont
  {Stoumpos}, \citenamefont {Cao}, \citenamefont {Clark}, \citenamefont
  {Young}, \citenamefont {Rondinelli}, \citenamefont {Jang}, \citenamefont
  {Hupp},\ and\ \citenamefont {Kanatzidis}}]{Stoumpos2016}%
  \BibitemOpen
  \bibfield  {author} {\bibinfo {author} {\bibfnamefont {C.~C.}\ \bibnamefont
  {Stoumpos}}, \bibinfo {author} {\bibfnamefont {D.~H.}\ \bibnamefont {Cao}},
  \bibinfo {author} {\bibfnamefont {D.~J.}\ \bibnamefont {Clark}}, \bibinfo
  {author} {\bibfnamefont {J.}~\bibnamefont {Young}}, \bibinfo {author}
  {\bibfnamefont {J.~M.}\ \bibnamefont {Rondinelli}}, \bibinfo {author}
  {\bibfnamefont {J.~I.}\ \bibnamefont {Jang}}, \bibinfo {author}
  {\bibfnamefont {J.~T.}\ \bibnamefont {Hupp}},\ and\ \bibinfo {author}
  {\bibfnamefont {M.~G.}\ \bibnamefont {Kanatzidis}},\ }\bibfield  {title}
  {\bibinfo {title} {Ruddlesden--popper hybrid lead iodide perovskite 2d
  homologous semiconductors},\ }\href
  {https://doi.org/10.1021/acs.chemmater.6b00847} {\bibfield  {journal}
  {\bibinfo  {journal} {Chemistry of Materials}\ }\textbf {\bibinfo {volume}
  {28}},\ \bibinfo {pages} {2852} (\bibinfo {year} {2016})}\BibitemShut
  {NoStop}%
\bibitem [{\citenamefont {Mitzi}(1996)}]{Mitzi1996}%
  \BibitemOpen
  \bibfield  {author} {\bibinfo {author} {\bibfnamefont {D.~B.}\ \bibnamefont
  {Mitzi}},\ }\bibfield  {title} {\bibinfo {title} {Synthesis, crystal
  structure, and optical and thermal properties of (c4h9nh3)2mi4 (m = ge, sn,
  pb)},\ }\href {https://doi.org/10.1021/cm9505097} {\bibfield  {journal}
  {\bibinfo  {journal} {Chemistry of Materials}\ }\textbf {\bibinfo {volume}
  {8}},\ \bibinfo {pages} {791} (\bibinfo {year} {1996})}\BibitemShut {NoStop}%
\bibitem [{\citenamefont {Wu}\ \emph {et~al.}(2015)\citenamefont {Wu},
  \citenamefont {Trinh},\ and\ \citenamefont {Zhu}}]{Wu2015}%
  \BibitemOpen
  \bibfield  {author} {\bibinfo {author} {\bibfnamefont {X.}~\bibnamefont
  {Wu}}, \bibinfo {author} {\bibfnamefont {M.~T.}\ \bibnamefont {Trinh}},\ and\
  \bibinfo {author} {\bibfnamefont {X.-Y.}\ \bibnamefont {Zhu}},\ }\bibfield
  {title} {\bibinfo {title} {Excitonic many-body interactions in
  two-dimensional lead iodide perovskite quantum wells},\ }\href
  {https://doi.org/10.1021/acs.jpcc.5b00148} {\bibfield  {journal} {\bibinfo
  {journal} {The Journal of Physical Chemistry C}\ }\textbf {\bibinfo {volume}
  {119}},\ \bibinfo {pages} {14714} (\bibinfo {year} {2015})}\BibitemShut
  {NoStop}%
\bibitem [{\citenamefont {Ziegler}\ \emph {et~al.}(2023)\citenamefont
  {Ziegler}, \citenamefont {Cho}, \citenamefont {Terres}, \citenamefont
  {Menahem}, \citenamefont {Taniguchi}, \citenamefont {Watanabe}, \citenamefont
  {Yaffe}, \citenamefont {Berkelbach},\ and\ \citenamefont
  {Chernikov}}]{https://doi.org/10.1002/adma.202210221}%
  \BibitemOpen
  \bibfield  {author} {\bibinfo {author} {\bibfnamefont {J.~D.}\ \bibnamefont
  {Ziegler}}, \bibinfo {author} {\bibfnamefont {Y.}~\bibnamefont {Cho}},
  \bibinfo {author} {\bibfnamefont {S.}~\bibnamefont {Terres}}, \bibinfo
  {author} {\bibfnamefont {M.}~\bibnamefont {Menahem}}, \bibinfo {author}
  {\bibfnamefont {T.}~\bibnamefont {Taniguchi}}, \bibinfo {author}
  {\bibfnamefont {K.}~\bibnamefont {Watanabe}}, \bibinfo {author}
  {\bibfnamefont {O.}~\bibnamefont {Yaffe}}, \bibinfo {author} {\bibfnamefont
  {T.~C.}\ \bibnamefont {Berkelbach}},\ and\ \bibinfo {author} {\bibfnamefont
  {A.}~\bibnamefont {Chernikov}},\ }\bibfield  {title} {\bibinfo {title}
  {Mobile trions in electrically tunable 2d hybrid perovskites},\ }\href
  {https://doi.org/https://doi.org/10.1002/adma.202210221} {\bibfield
  {journal} {\bibinfo  {journal} {Advanced Materials}\ }\textbf {\bibinfo
  {volume} {35}},\ \bibinfo {pages} {2210221} (\bibinfo {year}
  {2023})}\BibitemShut {NoStop}%
\bibitem [{\citenamefont {Blancon}\ \emph {et~al.}(2017)\citenamefont
  {Blancon}, \citenamefont {Tsai}, \citenamefont {Nie}, \citenamefont
  {Stoumpos}, \citenamefont {Pedesseau}, \citenamefont {Katan}, \citenamefont
  {Kepenekian}, \citenamefont {Soe}, \citenamefont {Appavoo}, \citenamefont
  {Sfeir}, \citenamefont {Tretiak}, \citenamefont {Ajayan}, \citenamefont
  {Kanatzidis}, \citenamefont {Even}, \citenamefont {Crochet},\ and\
  \citenamefont {Mohite}}]{doi:10.1126/science.aal4211}%
  \BibitemOpen
  \bibfield  {author} {\bibinfo {author} {\bibfnamefont {J.-C.}\ \bibnamefont
  {Blancon}}, \bibinfo {author} {\bibfnamefont {H.}~\bibnamefont {Tsai}},
  \bibinfo {author} {\bibfnamefont {W.}~\bibnamefont {Nie}}, \bibinfo {author}
  {\bibfnamefont {C.~C.}\ \bibnamefont {Stoumpos}}, \bibinfo {author}
  {\bibfnamefont {L.}~\bibnamefont {Pedesseau}}, \bibinfo {author}
  {\bibfnamefont {C.}~\bibnamefont {Katan}}, \bibinfo {author} {\bibfnamefont
  {M.}~\bibnamefont {Kepenekian}}, \bibinfo {author} {\bibfnamefont {C.~M.~M.}\
  \bibnamefont {Soe}}, \bibinfo {author} {\bibfnamefont {K.}~\bibnamefont
  {Appavoo}}, \bibinfo {author} {\bibfnamefont {M.~Y.}\ \bibnamefont {Sfeir}},
  \bibinfo {author} {\bibfnamefont {S.}~\bibnamefont {Tretiak}}, \bibinfo
  {author} {\bibfnamefont {P.~M.}\ \bibnamefont {Ajayan}}, \bibinfo {author}
  {\bibfnamefont {M.~G.}\ \bibnamefont {Kanatzidis}}, \bibinfo {author}
  {\bibfnamefont {J.}~\bibnamefont {Even}}, \bibinfo {author} {\bibfnamefont
  {J.~J.}\ \bibnamefont {Crochet}},\ and\ \bibinfo {author} {\bibfnamefont
  {A.~D.}\ \bibnamefont {Mohite}},\ }\bibfield  {title} {\bibinfo {title}
  {Extremely efficient internal exciton dissociation through edge states in
  layered 2d perovskites},\ }\href {https://doi.org/10.1126/science.aal4211}
  {\bibfield  {journal} {\bibinfo  {journal} {Science}\ }\textbf {\bibinfo
  {volume} {355}},\ \bibinfo {pages} {1288} (\bibinfo {year}
  {2017})}\BibitemShut {NoStop}%
\bibitem [{\citenamefont {Qin}\ \emph {et~al.}(2022)\citenamefont {Qin},
  \citenamefont {Li}, \citenamefont {Gao}, \citenamefont {Chen}, \citenamefont
  {Li}, \citenamefont {Xu}, \citenamefont {Li}, \citenamefont {Jiang},
  \citenamefont {Li}, \citenamefont {Wu}, \citenamefont {Quan}, \citenamefont
  {Ye}, \citenamefont {Zhang}, \citenamefont {Lin}, \citenamefont {Pedesseau},
  \citenamefont {Even}, \citenamefont {Lu},\ and\ \citenamefont
  {Bu}}]{https://doi.org/10.1002/adma.202201666}%
  \BibitemOpen
  \bibfield  {author} {\bibinfo {author} {\bibfnamefont {Y.}~\bibnamefont
  {Qin}}, \bibinfo {author} {\bibfnamefont {Z.-G.}\ \bibnamefont {Li}},
  \bibinfo {author} {\bibfnamefont {F.-F.}\ \bibnamefont {Gao}}, \bibinfo
  {author} {\bibfnamefont {H.}~\bibnamefont {Chen}}, \bibinfo {author}
  {\bibfnamefont {X.}~\bibnamefont {Li}}, \bibinfo {author} {\bibfnamefont
  {B.}~\bibnamefont {Xu}}, \bibinfo {author} {\bibfnamefont {Q.}~\bibnamefont
  {Li}}, \bibinfo {author} {\bibfnamefont {X.}~\bibnamefont {Jiang}}, \bibinfo
  {author} {\bibfnamefont {W.}~\bibnamefont {Li}}, \bibinfo {author}
  {\bibfnamefont {X.}~\bibnamefont {Wu}}, \bibinfo {author} {\bibfnamefont
  {Z.}~\bibnamefont {Quan}}, \bibinfo {author} {\bibfnamefont {L.}~\bibnamefont
  {Ye}}, \bibinfo {author} {\bibfnamefont {Y.}~\bibnamefont {Zhang}}, \bibinfo
  {author} {\bibfnamefont {Z.}~\bibnamefont {Lin}}, \bibinfo {author}
  {\bibfnamefont {L.}~\bibnamefont {Pedesseau}}, \bibinfo {author}
  {\bibfnamefont {J.}~\bibnamefont {Even}}, \bibinfo {author} {\bibfnamefont
  {P.}~\bibnamefont {Lu}},\ and\ \bibinfo {author} {\bibfnamefont {X.-H.}\
  \bibnamefont {Bu}},\ }\bibfield  {title} {\bibinfo {title} {Dangling
  octahedra enable edge states in 2d lead halide perovskites},\ }\href
  {https://doi.org/https://doi.org/10.1002/adma.202201666} {\bibfield
  {journal} {\bibinfo  {journal} {Advanced Materials}\ }\textbf {\bibinfo
  {volume} {34}},\ \bibinfo {pages} {2201666} (\bibinfo {year}
  {2022})}\BibitemShut {NoStop}%
\bibitem [{\citenamefont {Zhao}\ \emph {et~al.}(2019)\citenamefont {Zhao},
  \citenamefont {Tian}, \citenamefont {Leng}, \citenamefont {Zhao},\ and\
  \citenamefont {Jin}}]{Zhao2019}%
  \BibitemOpen
  \bibfield  {author} {\bibinfo {author} {\bibfnamefont {C.}~\bibnamefont
  {Zhao}}, \bibinfo {author} {\bibfnamefont {W.}~\bibnamefont {Tian}}, \bibinfo
  {author} {\bibfnamefont {J.}~\bibnamefont {Leng}}, \bibinfo {author}
  {\bibfnamefont {Y.}~\bibnamefont {Zhao}},\ and\ \bibinfo {author}
  {\bibfnamefont {S.}~\bibnamefont {Jin}},\ }\bibfield  {title} {\bibinfo
  {title} {Controlling the property of edges in layered 2d perovskite single
  crystals},\ }\href {https://doi.org/10.1021/acs.jpclett.9b01193} {\bibfield
  {journal} {\bibinfo  {journal} {The Journal of Physical Chemistry Letters}\
  }\textbf {\bibinfo {volume} {10}},\ \bibinfo {pages} {3950} (\bibinfo {year}
  {2019})}\BibitemShut {NoStop}%
\bibitem [{\citenamefont {Zhang}\ \emph {et~al.}(2019)\citenamefont {Zhang},
  \citenamefont {Fang}, \citenamefont {Long},\ and\ \citenamefont
  {Prezhdo}}]{Zhang2019}%
  \BibitemOpen
  \bibfield  {author} {\bibinfo {author} {\bibfnamefont {Z.}~\bibnamefont
  {Zhang}}, \bibinfo {author} {\bibfnamefont {W.-H.}\ \bibnamefont {Fang}},
  \bibinfo {author} {\bibfnamefont {R.}~\bibnamefont {Long}},\ and\ \bibinfo
  {author} {\bibfnamefont {O.~V.}\ \bibnamefont {Prezhdo}},\ }\bibfield
  {title} {\bibinfo {title} {Exciton dissociation and suppressed charge
  recombination at 2d perovskite edges: Key roles of unsaturated halide bonds
  and thermal disorder},\ }\href {https://doi.org/10.1021/jacs.9b06046}
  {\bibfield  {journal} {\bibinfo  {journal} {Journal of the American Chemical
  Society}\ }\textbf {\bibinfo {volume} {141}},\ \bibinfo {pages} {15557}
  (\bibinfo {year} {2019})}\BibitemShut {NoStop}%
\bibitem [{\citenamefont {Srimath~Kandada}\ and\ \citenamefont
  {Silva}(2020)}]{SrimathKandada2020}%
  \BibitemOpen
  \bibfield  {author} {\bibinfo {author} {\bibfnamefont {A.~R.}\ \bibnamefont
  {Srimath~Kandada}}\ and\ \bibinfo {author} {\bibfnamefont {C.}~\bibnamefont
  {Silva}},\ }\bibfield  {title} {\bibinfo {title} {Exciton polarons in
  two-dimensional hybrid metal-halide perovskites},\ }\href
  {https://doi.org/10.1021/acs.jpclett.9b02342} {\bibfield  {journal} {\bibinfo
   {journal} {The Journal of Physical Chemistry Letters}\ }\textbf {\bibinfo
  {volume} {11}},\ \bibinfo {pages} {3173} (\bibinfo {year}
  {2020})}\BibitemShut {NoStop}%
\bibitem [{\citenamefont {Yin}\ \emph {et~al.}(2017)\citenamefont {Yin},
  \citenamefont {Li}, \citenamefont {Cortecchia}, \citenamefont {Soci},\ and\
  \citenamefont {Br{\'e}das}}]{Yin2017}%
  \BibitemOpen
  \bibfield  {author} {\bibinfo {author} {\bibfnamefont {J.}~\bibnamefont
  {Yin}}, \bibinfo {author} {\bibfnamefont {H.}~\bibnamefont {Li}}, \bibinfo
  {author} {\bibfnamefont {D.}~\bibnamefont {Cortecchia}}, \bibinfo {author}
  {\bibfnamefont {C.}~\bibnamefont {Soci}},\ and\ \bibinfo {author}
  {\bibfnamefont {J.-L.}\ \bibnamefont {Br{\'e}das}},\ }\bibfield  {title}
  {\bibinfo {title} {Excitonic and polaronic properties of 2d hybrid
  organic--inorganic perovskites},\ }\href
  {https://doi.org/10.1021/acsenergylett.6b00659} {\bibfield  {journal}
  {\bibinfo  {journal} {ACS Energy Letters}\ }\textbf {\bibinfo {volume} {2}},\
  \bibinfo {pages} {417} (\bibinfo {year} {2017})}\BibitemShut {NoStop}%
\bibitem [{\citenamefont {Simbula}\ \emph {et~al.}(2021)\citenamefont
  {Simbula}, \citenamefont {Pau}, \citenamefont {Wang}, \citenamefont {Liu},
  \citenamefont {Sarritzu}, \citenamefont {Lai}, \citenamefont {Lodde},
  \citenamefont {Mattana}, \citenamefont {Mula}, \citenamefont {Geddo~Lehmann},
  \citenamefont {Spanopoulos}, \citenamefont {Kanatzidis}, \citenamefont
  {Marongiu}, \citenamefont {Quochi}, \citenamefont {Saba}, \citenamefont
  {Mura},\ and\ \citenamefont
  {Bongiovanni}}]{https://doi.org/10.1002/adom.202100295}%
  \BibitemOpen
  \bibfield  {author} {\bibinfo {author} {\bibfnamefont {A.}~\bibnamefont
  {Simbula}}, \bibinfo {author} {\bibfnamefont {R.}~\bibnamefont {Pau}},
  \bibinfo {author} {\bibfnamefont {Q.}~\bibnamefont {Wang}}, \bibinfo {author}
  {\bibfnamefont {F.}~\bibnamefont {Liu}}, \bibinfo {author} {\bibfnamefont
  {V.}~\bibnamefont {Sarritzu}}, \bibinfo {author} {\bibfnamefont
  {S.}~\bibnamefont {Lai}}, \bibinfo {author} {\bibfnamefont {M.}~\bibnamefont
  {Lodde}}, \bibinfo {author} {\bibfnamefont {F.}~\bibnamefont {Mattana}},
  \bibinfo {author} {\bibfnamefont {G.}~\bibnamefont {Mula}}, \bibinfo {author}
  {\bibfnamefont {A.}~\bibnamefont {Geddo~Lehmann}}, \bibinfo {author}
  {\bibfnamefont {I.~D.}\ \bibnamefont {Spanopoulos}}, \bibinfo {author}
  {\bibfnamefont {M.~G.}\ \bibnamefont {Kanatzidis}}, \bibinfo {author}
  {\bibfnamefont {D.}~\bibnamefont {Marongiu}}, \bibinfo {author}
  {\bibfnamefont {F.}~\bibnamefont {Quochi}}, \bibinfo {author} {\bibfnamefont
  {M.}~\bibnamefont {Saba}}, \bibinfo {author} {\bibfnamefont {A.}~\bibnamefont
  {Mura}},\ and\ \bibinfo {author} {\bibfnamefont {G.}~\bibnamefont
  {Bongiovanni}},\ }\bibfield  {title} {\bibinfo {title} {Polaron plasma in
  equilibrium with bright excitons in 2d and 3d hybrid perovskites},\ }\href
  {https://doi.org/https://doi.org/10.1002/adom.202100295} {\bibfield
  {journal} {\bibinfo  {journal} {Advanced Optical Materials}\ }\textbf
  {\bibinfo {volume} {9}},\ \bibinfo {pages} {2100295} (\bibinfo {year}
  {2021})}\BibitemShut {NoStop}%
\bibitem [{\citenamefont {Guzelturk}\ \emph {et~al.}(2021)\citenamefont
  {Guzelturk}, \citenamefont {Winkler}, \citenamefont {Van~de Goor},
  \citenamefont {Smith}, \citenamefont {Bourelle}, \citenamefont {Feldmann},
  \citenamefont {Trigo}, \citenamefont {Teitelbaum}, \citenamefont
  {Steinr{\"u}ck}, \citenamefont {de~la Pena}, \citenamefont {Alonso-Mori},
  \citenamefont {Zhu}, \citenamefont {Sato}, \citenamefont {Karunadasa},
  \citenamefont {Toney}, \citenamefont {Deschler},\ and\ \citenamefont
  {Lindenberg}}]{Guzelturk2021}%
  \BibitemOpen
  \bibfield  {author} {\bibinfo {author} {\bibfnamefont {B.}~\bibnamefont
  {Guzelturk}}, \bibinfo {author} {\bibfnamefont {T.}~\bibnamefont {Winkler}},
  \bibinfo {author} {\bibfnamefont {T.~W.~J.}\ \bibnamefont {Van~de Goor}},
  \bibinfo {author} {\bibfnamefont {M.~D.}\ \bibnamefont {Smith}}, \bibinfo
  {author} {\bibfnamefont {S.~A.}\ \bibnamefont {Bourelle}}, \bibinfo {author}
  {\bibfnamefont {S.}~\bibnamefont {Feldmann}}, \bibinfo {author}
  {\bibfnamefont {M.}~\bibnamefont {Trigo}}, \bibinfo {author} {\bibfnamefont
  {S.~W.}\ \bibnamefont {Teitelbaum}}, \bibinfo {author} {\bibfnamefont
  {H.-G.}\ \bibnamefont {Steinr{\"u}ck}}, \bibinfo {author} {\bibfnamefont
  {G.~A.}\ \bibnamefont {de~la Pena}}, \bibinfo {author} {\bibfnamefont
  {R.}~\bibnamefont {Alonso-Mori}}, \bibinfo {author} {\bibfnamefont
  {D.}~\bibnamefont {Zhu}}, \bibinfo {author} {\bibfnamefont {T.}~\bibnamefont
  {Sato}}, \bibinfo {author} {\bibfnamefont {H.~I.}\ \bibnamefont
  {Karunadasa}}, \bibinfo {author} {\bibfnamefont {M.~F.}\ \bibnamefont
  {Toney}}, \bibinfo {author} {\bibfnamefont {F.}~\bibnamefont {Deschler}},\
  and\ \bibinfo {author} {\bibfnamefont {A.~M.}\ \bibnamefont {Lindenberg}},\
  }\bibfield  {title} {\bibinfo {title} {Visualization of dynamic polaronic
  strain fields in hybrid lead halide perovskites},\ }\href
  {https://doi.org/10.1038/s41563-020-00865-5} {\bibfield  {journal} {\bibinfo
  {journal} {Nature Materials}\ }\textbf {\bibinfo {volume} {20}},\ \bibinfo
  {pages} {618} (\bibinfo {year} {2021})}\BibitemShut {NoStop}%
\bibitem [{\citenamefont {Zheng}\ \emph {et~al.}(2016)\citenamefont {Zheng},
  \citenamefont {Abdellah}, \citenamefont {Zhu}, \citenamefont {Kong},
  \citenamefont {Jennings}, \citenamefont {Kurtz}, \citenamefont {Messing},
  \citenamefont {Niu}, \citenamefont {Gosztola}, \citenamefont {Al-Marri},
  \citenamefont {Zhang}, \citenamefont {Pullerits},\ and\ \citenamefont
  {Canton}}]{Zheng2016}%
  \BibitemOpen
  \bibfield  {author} {\bibinfo {author} {\bibfnamefont {K.}~\bibnamefont
  {Zheng}}, \bibinfo {author} {\bibfnamefont {M.}~\bibnamefont {Abdellah}},
  \bibinfo {author} {\bibfnamefont {Q.}~\bibnamefont {Zhu}}, \bibinfo {author}
  {\bibfnamefont {Q.}~\bibnamefont {Kong}}, \bibinfo {author} {\bibfnamefont
  {G.}~\bibnamefont {Jennings}}, \bibinfo {author} {\bibfnamefont {C.~A.}\
  \bibnamefont {Kurtz}}, \bibinfo {author} {\bibfnamefont {M.~E.}\ \bibnamefont
  {Messing}}, \bibinfo {author} {\bibfnamefont {Y.}~\bibnamefont {Niu}},
  \bibinfo {author} {\bibfnamefont {D.~J.}\ \bibnamefont {Gosztola}}, \bibinfo
  {author} {\bibfnamefont {M.~J.}\ \bibnamefont {Al-Marri}}, \bibinfo {author}
  {\bibfnamefont {X.}~\bibnamefont {Zhang}}, \bibinfo {author} {\bibfnamefont
  {T.}~\bibnamefont {Pullerits}},\ and\ \bibinfo {author} {\bibfnamefont
  {S.~E.}\ \bibnamefont {Canton}},\ }\bibfield  {title} {\bibinfo {title}
  {Direct experimental evidence for photoinduced strong-coupling polarons in
  organolead halide perovskite nanoparticles},\ }\href
  {https://doi.org/10.1021/acs.jpclett.6b02046} {\bibfield  {journal} {\bibinfo
   {journal} {The Journal of Physical Chemistry Letters}\ }\textbf {\bibinfo
  {volume} {7}},\ \bibinfo {pages} {4535} (\bibinfo {year} {2016})}\BibitemShut
  {NoStop}%
\bibitem [{\citenamefont {Liu}\ \emph {et~al.}(2020)\citenamefont {Liu},
  \citenamefont {Tsai}, \citenamefont {Nie}, \citenamefont {Gosztola},\ and\
  \citenamefont {Zhang}}]{Liu2020}%
  \BibitemOpen
  \bibfield  {author} {\bibinfo {author} {\bibfnamefont {C.}~\bibnamefont
  {Liu}}, \bibinfo {author} {\bibfnamefont {H.}~\bibnamefont {Tsai}}, \bibinfo
  {author} {\bibfnamefont {W.}~\bibnamefont {Nie}}, \bibinfo {author}
  {\bibfnamefont {D.~J.}\ \bibnamefont {Gosztola}},\ and\ \bibinfo {author}
  {\bibfnamefont {X.}~\bibnamefont {Zhang}},\ }\bibfield  {title} {\bibinfo
  {title} {Direct spectroscopic observation of the hole polaron in lead halide
  perovskites},\ }\href {https://doi.org/10.1021/acs.jpclett.0c01708}
  {\bibfield  {journal} {\bibinfo  {journal} {The Journal of Physical Chemistry
  Letters}\ }\textbf {\bibinfo {volume} {11}},\ \bibinfo {pages} {6256}
  (\bibinfo {year} {2020})}\BibitemShut {NoStop}%
\bibitem [{\citenamefont {Buizza}\ and\ \citenamefont
  {Herz}(2021)}]{https://doi.org/10.1002/adma.202007057}%
  \BibitemOpen
  \bibfield  {author} {\bibinfo {author} {\bibfnamefont {L.~R.~V.}\
  \bibnamefont {Buizza}}\ and\ \bibinfo {author} {\bibfnamefont {L.~M.}\
  \bibnamefont {Herz}},\ }\bibfield  {title} {\bibinfo {title} {Polarons and
  charge localization in metal-halide semiconductors for photovoltaic and
  light-emitting devices},\ }\href
  {https://doi.org/https://doi.org/10.1002/adma.202007057} {\bibfield
  {journal} {\bibinfo  {journal} {Advanced Materials}\ }\textbf {\bibinfo
  {volume} {33}},\ \bibinfo {pages} {2007057} (\bibinfo {year}
  {2021})}\BibitemShut {NoStop}%
\bibitem [{\citenamefont {Tao}\ \emph {et~al.}(2022)\citenamefont {Tao},
  \citenamefont {Zhang},\ and\ \citenamefont {Zhu}}]{Tao2022}%
  \BibitemOpen
  \bibfield  {author} {\bibinfo {author} {\bibfnamefont {W.}~\bibnamefont
  {Tao}}, \bibinfo {author} {\bibfnamefont {Y.}~\bibnamefont {Zhang}},\ and\
  \bibinfo {author} {\bibfnamefont {H.}~\bibnamefont {Zhu}},\ }\bibfield
  {title} {\bibinfo {title} {Dynamic exciton polaron in two-dimensional lead
  halide perovskites and implications for optoelectronic applications},\ }\href
  {https://doi.org/10.1021/acs.accounts.1c00626} {\bibfield  {journal}
  {\bibinfo  {journal} {Accounts of Chemical Research}\ }\textbf {\bibinfo
  {volume} {55}},\ \bibinfo {pages} {345} (\bibinfo {year} {2022})}\BibitemShut
  {NoStop}%
\bibitem [{\citenamefont {Franchini}\ \emph {et~al.}(2021)\citenamefont
  {Franchini}, \citenamefont {Reticcioli}, \citenamefont {Setvin},\ and\
  \citenamefont {Diebold}}]{Franchini2021}%
  \BibitemOpen
  \bibfield  {author} {\bibinfo {author} {\bibfnamefont {C.}~\bibnamefont
  {Franchini}}, \bibinfo {author} {\bibfnamefont {M.}~\bibnamefont
  {Reticcioli}}, \bibinfo {author} {\bibfnamefont {M.}~\bibnamefont {Setvin}},\
  and\ \bibinfo {author} {\bibfnamefont {U.}~\bibnamefont {Diebold}},\
  }\bibfield  {title} {\bibinfo {title} {Polarons in materials},\ }\href
  {https://doi.org/10.1038/s41578-021-00289-w} {\bibfield  {journal} {\bibinfo
  {journal} {Nature Reviews Materials}\ }\textbf {\bibinfo {volume} {6}},\
  \bibinfo {pages} {560} (\bibinfo {year} {2021})}\BibitemShut {NoStop}%
\bibitem [{\citenamefont {Simbula}\ \emph {et~al.}(2023)\citenamefont
  {Simbula}, \citenamefont {Wu}, \citenamefont {Pitzalis}, \citenamefont {Pau},
  \citenamefont {Lai}, \citenamefont {Liu}, \citenamefont {Matta},
  \citenamefont {Marongiu}, \citenamefont {Quochi}, \citenamefont {Saba},
  \citenamefont {Mura},\ and\ \citenamefont {Bongiovanni}}]{Simbula2023}%
  \BibitemOpen
  \bibfield  {author} {\bibinfo {author} {\bibfnamefont {A.}~\bibnamefont
  {Simbula}}, \bibinfo {author} {\bibfnamefont {L.}~\bibnamefont {Wu}},
  \bibinfo {author} {\bibfnamefont {F.}~\bibnamefont {Pitzalis}}, \bibinfo
  {author} {\bibfnamefont {R.}~\bibnamefont {Pau}}, \bibinfo {author}
  {\bibfnamefont {S.}~\bibnamefont {Lai}}, \bibinfo {author} {\bibfnamefont
  {F.}~\bibnamefont {Liu}}, \bibinfo {author} {\bibfnamefont {S.}~\bibnamefont
  {Matta}}, \bibinfo {author} {\bibfnamefont {D.}~\bibnamefont {Marongiu}},
  \bibinfo {author} {\bibfnamefont {F.}~\bibnamefont {Quochi}}, \bibinfo
  {author} {\bibfnamefont {M.}~\bibnamefont {Saba}}, \bibinfo {author}
  {\bibfnamefont {A.}~\bibnamefont {Mura}},\ and\ \bibinfo {author}
  {\bibfnamefont {G.}~\bibnamefont {Bongiovanni}},\ }\bibfield  {title}
  {\bibinfo {title} {Exciton dissociation in 2d layered metal-halide
  perovskites},\ }\href {https://doi.org/10.1038/s41467-023-39831-5} {\bibfield
   {journal} {\bibinfo  {journal} {Nature Communications}\ }\textbf {\bibinfo
  {volume} {14}},\ \bibinfo {pages} {4125} (\bibinfo {year}
  {2023})}\BibitemShut {NoStop}%
\bibitem [{\citenamefont {Motti}\ \emph {et~al.}(2023)\citenamefont {Motti},
  \citenamefont {Kober-Czerny}, \citenamefont {Righetto}, \citenamefont
  {Holzhey}, \citenamefont {Smith}, \citenamefont {Kraus}, \citenamefont
  {Snaith}, \citenamefont {Johnston},\ and\ \citenamefont
  {Herz}}]{https://doi.org/10.1002/adfm.202300363}%
  \BibitemOpen
  \bibfield  {author} {\bibinfo {author} {\bibfnamefont {S.~G.}\ \bibnamefont
  {Motti}}, \bibinfo {author} {\bibfnamefont {M.}~\bibnamefont {Kober-Czerny}},
  \bibinfo {author} {\bibfnamefont {M.}~\bibnamefont {Righetto}}, \bibinfo
  {author} {\bibfnamefont {P.}~\bibnamefont {Holzhey}}, \bibinfo {author}
  {\bibfnamefont {J.}~\bibnamefont {Smith}}, \bibinfo {author} {\bibfnamefont
  {H.}~\bibnamefont {Kraus}}, \bibinfo {author} {\bibfnamefont {H.~J.}\
  \bibnamefont {Snaith}}, \bibinfo {author} {\bibfnamefont {M.~B.}\
  \bibnamefont {Johnston}},\ and\ \bibinfo {author} {\bibfnamefont {L.~M.}\
  \bibnamefont {Herz}},\ }\bibfield  {title} {\bibinfo {title} {Exciton
  formation dynamics and band-like free charge-carrier transport in 2d metal
  halide perovskite semiconductors},\ }\href
  {https://doi.org/https://doi.org/10.1002/adfm.202300363} {\bibfield
  {journal} {\bibinfo  {journal} {Advanced Functional Materials}\ }\textbf
  {\bibinfo {volume} {33}},\ \bibinfo {pages} {2300363} (\bibinfo {year}
  {2023})}\BibitemShut {NoStop}%
\bibitem [{\citenamefont {Thouin}\ \emph {et~al.}(2019)\citenamefont {Thouin},
  \citenamefont {Valverde-Ch{\'a}vez}, \citenamefont {Quarti}, \citenamefont
  {Cortecchia}, \citenamefont {Bargigia}, \citenamefont {Beljonne},
  \citenamefont {Petrozza}, \citenamefont {Silva},\ and\ \citenamefont
  {Srimath~Kandada}}]{Thouin2019}%
  \BibitemOpen
  \bibfield  {author} {\bibinfo {author} {\bibfnamefont {F.}~\bibnamefont
  {Thouin}}, \bibinfo {author} {\bibfnamefont {D.~A.}\ \bibnamefont
  {Valverde-Ch{\'a}vez}}, \bibinfo {author} {\bibfnamefont {C.}~\bibnamefont
  {Quarti}}, \bibinfo {author} {\bibfnamefont {D.}~\bibnamefont {Cortecchia}},
  \bibinfo {author} {\bibfnamefont {I.}~\bibnamefont {Bargigia}}, \bibinfo
  {author} {\bibfnamefont {D.}~\bibnamefont {Beljonne}}, \bibinfo {author}
  {\bibfnamefont {A.}~\bibnamefont {Petrozza}}, \bibinfo {author}
  {\bibfnamefont {C.}~\bibnamefont {Silva}},\ and\ \bibinfo {author}
  {\bibfnamefont {A.~R.}\ \bibnamefont {Srimath~Kandada}},\ }\bibfield  {title}
  {\bibinfo {title} {Phonon coherences reveal the polaronic character of
  excitons in two-dimensional lead halide perovskites},\ }\href
  {https://doi.org/10.1038/s41563-018-0262-7} {\bibfield  {journal} {\bibinfo
  {journal} {Nature Materials}\ }\textbf {\bibinfo {volume} {18}},\ \bibinfo
  {pages} {349} (\bibinfo {year} {2019})}\BibitemShut {NoStop}%
\bibitem [{\citenamefont {Das}\ \emph {et~al.}(2015)\citenamefont {Das},
  \citenamefont {Coulter},\ and\ \citenamefont
  {Manousakis}}]{PhysRevB.91.115105}%
  \BibitemOpen
  \bibfield  {author} {\bibinfo {author} {\bibfnamefont {S.}~\bibnamefont
  {Das}}, \bibinfo {author} {\bibfnamefont {J.~E.}\ \bibnamefont {Coulter}},\
  and\ \bibinfo {author} {\bibfnamefont {E.}~\bibnamefont {Manousakis}},\
  }\bibfield  {title} {\bibinfo {title} {Convergence of quasiparticle
  self-consistent $gw$ calculations of transition-metal monoxides},\ }\href
  {https://doi.org/10.1103/PhysRevB.91.115105} {\bibfield  {journal} {\bibinfo
  {journal} {Phys. Rev. B}\ }\textbf {\bibinfo {volume} {91}},\ \bibinfo
  {pages} {115105} (\bibinfo {year} {2015})}\BibitemShut {NoStop}%
\bibitem [{\citenamefont {Coulter}\ \emph {et~al.}(2014)\citenamefont
  {Coulter}, \citenamefont {Manousakis},\ and\ \citenamefont
  {Gali}}]{PhysRevB.90.165142}%
  \BibitemOpen
  \bibfield  {author} {\bibinfo {author} {\bibfnamefont {J.~E.}\ \bibnamefont
  {Coulter}}, \bibinfo {author} {\bibfnamefont {E.}~\bibnamefont
  {Manousakis}},\ and\ \bibinfo {author} {\bibfnamefont {A.}~\bibnamefont
  {Gali}},\ }\bibfield  {title} {\bibinfo {title} {Optoelectronic excitations
  and photovoltaic effect in strongly correlated materials},\ }\href
  {https://doi.org/10.1103/PhysRevB.90.165142} {\bibfield  {journal} {\bibinfo
  {journal} {Phys. Rev. B}\ }\textbf {\bibinfo {volume} {90}},\ \bibinfo
  {pages} {165142} (\bibinfo {year} {2014})}\BibitemShut {NoStop}%
\bibitem [{\citenamefont {van Schilfgaarde}\ \emph {et~al.}(2006)\citenamefont
  {van Schilfgaarde}, \citenamefont {Kotani},\ and\ \citenamefont
  {Faleev}}]{PhysRevLett.96.226402}%
  \BibitemOpen
  \bibfield  {author} {\bibinfo {author} {\bibfnamefont {M.}~\bibnamefont {van
  Schilfgaarde}}, \bibinfo {author} {\bibfnamefont {T.}~\bibnamefont
  {Kotani}},\ and\ \bibinfo {author} {\bibfnamefont {S.}~\bibnamefont
  {Faleev}},\ }\bibfield  {title} {\bibinfo {title} {Quasiparticle
  self-consistent $gw$ theory},\ }\href
  {https://doi.org/10.1103/PhysRevLett.96.226402} {\bibfield  {journal}
  {\bibinfo  {journal} {Phys. Rev. Lett.}\ }\textbf {\bibinfo {volume} {96}},\
  \bibinfo {pages} {226402} (\bibinfo {year} {2006})}\BibitemShut {NoStop}%
\bibitem [{\citenamefont {Shishkin}\ \emph {et~al.}(2007)\citenamefont
  {Shishkin}, \citenamefont {Marsman},\ and\ \citenamefont
  {Kresse}}]{PhysRevLett.99.246403}%
  \BibitemOpen
  \bibfield  {author} {\bibinfo {author} {\bibfnamefont {M.}~\bibnamefont
  {Shishkin}}, \bibinfo {author} {\bibfnamefont {M.}~\bibnamefont {Marsman}},\
  and\ \bibinfo {author} {\bibfnamefont {G.}~\bibnamefont {Kresse}},\
  }\bibfield  {title} {\bibinfo {title} {Accurate quasiparticle spectra from
  self-consistent gw calculations with vertex corrections},\ }\href
  {https://doi.org/10.1103/PhysRevLett.99.246403} {\bibfield  {journal}
  {\bibinfo  {journal} {Phys. Rev. Lett.}\ }\textbf {\bibinfo {volume} {99}},\
  \bibinfo {pages} {246403} (\bibinfo {year} {2007})}\BibitemShut {NoStop}%
\bibitem [{\citenamefont {deQuilettes}\ \emph {et~al.}(2019)\citenamefont
  {deQuilettes}, \citenamefont {Frohna}, \citenamefont {Emin}, \citenamefont
  {Kirchartz}, \citenamefont {Bulovic}, \citenamefont {Ginger},\ and\
  \citenamefont {Stranks}}]{deQuilettes2019}%
  \BibitemOpen
  \bibfield  {author} {\bibinfo {author} {\bibfnamefont {D.~W.}\ \bibnamefont
  {deQuilettes}}, \bibinfo {author} {\bibfnamefont {K.}~\bibnamefont {Frohna}},
  \bibinfo {author} {\bibfnamefont {D.}~\bibnamefont {Emin}}, \bibinfo {author}
  {\bibfnamefont {T.}~\bibnamefont {Kirchartz}}, \bibinfo {author}
  {\bibfnamefont {V.}~\bibnamefont {Bulovic}}, \bibinfo {author} {\bibfnamefont
  {D.~S.}\ \bibnamefont {Ginger}},\ and\ \bibinfo {author} {\bibfnamefont
  {S.~D.}\ \bibnamefont {Stranks}},\ }\bibfield  {title} {\bibinfo {title}
  {Charge-carrier recombination in halide perovskites},\ }\href
  {https://doi.org/10.1021/acs.chemrev.9b00169} {\bibfield  {journal} {\bibinfo
   {journal} {Chemical Reviews}\ }\textbf {\bibinfo {volume} {119}},\ \bibinfo
  {pages} {11007} (\bibinfo {year} {2019})}\BibitemShut {NoStop}%
\bibitem [{\citenamefont {Saha}(1920)}]{doi:10.1080/14786441008636148}%
  \BibitemOpen
  \bibfield  {author} {\bibinfo {author} {\bibfnamefont {M.~N.}\ \bibnamefont
  {Saha}},\ }\bibfield  {title} {\bibinfo {title} {Liii. ionization in the
  solar chromosphere},\ }\href {https://doi.org/10.1080/14786441008636148}
  {\bibfield  {journal} {\bibinfo  {journal} {The London, Edinburgh, and Dublin
  Philosophical Magazine and Journal of Science}\ }\textbf {\bibinfo {volume}
  {40}},\ \bibinfo {pages} {472} (\bibinfo {year} {1920})}\BibitemShut
  {NoStop}%
\bibitem [{\citenamefont {Saha}\ and\ \citenamefont
  {Fowler}(1921)}]{doi:10.1098/rspa.1921.0029}%
  \BibitemOpen
  \bibfield  {author} {\bibinfo {author} {\bibfnamefont {M.~N.}\ \bibnamefont
  {Saha}}\ and\ \bibinfo {author} {\bibfnamefont {A.}~\bibnamefont {Fowler}},\
  }\bibfield  {title} {\bibinfo {title} {On a physical theory of stellar
  spectra},\ }\href {https://doi.org/10.1098/rspa.1921.0029} {\bibfield
  {journal} {\bibinfo  {journal} {Proceedings of the Royal Society of London.
  Series A, Containing Papers of a Mathematical and Physical Character}\
  }\textbf {\bibinfo {volume} {99}},\ \bibinfo {pages} {135} (\bibinfo {year}
  {1921})}\BibitemShut {NoStop}%
\bibitem [{\citenamefont {Cho}\ \emph {et~al.}(2021)\citenamefont {Cho},
  \citenamefont {Greene},\ and\ \citenamefont
  {Berkelbach}}]{PhysRevLett.126.216402}%
  \BibitemOpen
  \bibfield  {author} {\bibinfo {author} {\bibfnamefont {Y.}~\bibnamefont
  {Cho}}, \bibinfo {author} {\bibfnamefont {S.~M.}\ \bibnamefont {Greene}},\
  and\ \bibinfo {author} {\bibfnamefont {T.~C.}\ \bibnamefont {Berkelbach}},\
  }\bibfield  {title} {\bibinfo {title} {Simulations of trions and biexcitons
  in layered hybrid organic-inorganic lead halide perovskites},\ }\href
  {https://doi.org/10.1103/PhysRevLett.126.216402} {\bibfield  {journal}
  {\bibinfo  {journal} {Phys. Rev. Lett.}\ }\textbf {\bibinfo {volume} {126}},\
  \bibinfo {pages} {216402} (\bibinfo {year} {2021})}\BibitemShut {NoStop}%
\bibitem [{\citenamefont {Kanemitsu}(2019)}]{10.1063/1.5125628}%
  \BibitemOpen
  \bibfield  {author} {\bibinfo {author} {\bibfnamefont {Y.}~\bibnamefont
  {Kanemitsu}},\ }\bibfield  {title} {\bibinfo {title} {{Trion dynamics in lead
  halide perovskite nanocrystals}},\ }\href {https://doi.org/10.1063/1.5125628}
  {\bibfield  {journal} {\bibinfo  {journal} {The Journal of Chemical Physics}\
  }\textbf {\bibinfo {volume} {151}},\ \bibinfo {pages} {170902} (\bibinfo
  {year} {2019})}\BibitemShut {NoStop}%
\bibitem [{\citenamefont {Brennan}\ \emph {et~al.}(2018)\citenamefont
  {Brennan}, \citenamefont {Draguta}, \citenamefont {Kamat},\ and\
  \citenamefont {Kuno}}]{Brennan2018}%
  \BibitemOpen
  \bibfield  {author} {\bibinfo {author} {\bibfnamefont {M.~C.}\ \bibnamefont
  {Brennan}}, \bibinfo {author} {\bibfnamefont {S.}~\bibnamefont {Draguta}},
  \bibinfo {author} {\bibfnamefont {P.~V.}\ \bibnamefont {Kamat}},\ and\
  \bibinfo {author} {\bibfnamefont {M.}~\bibnamefont {Kuno}},\ }\bibfield
  {title} {\bibinfo {title} {Light-induced anion phase segregation in mixed
  halide perovskites},\ }\href {https://doi.org/10.1021/acsenergylett.7b01151}
  {\bibfield  {journal} {\bibinfo  {journal} {ACS Energy Letters}\ }\textbf
  {\bibinfo {volume} {3}},\ \bibinfo {pages} {204} (\bibinfo {year}
  {2018})}\BibitemShut {NoStop}%
\bibitem [{\citenamefont {Brenes}\ \emph {et~al.}(2017)\citenamefont {Brenes},
  \citenamefont {Guo}, \citenamefont {Osherov}, \citenamefont {Noel},
  \citenamefont {Eames}, \citenamefont {Hutter}, \citenamefont {Pathak},
  \citenamefont {Niroui}, \citenamefont {Friend}, \citenamefont {Islam},
  \citenamefont {Snaith}, \citenamefont {Bulović}, \citenamefont {Savenije},\
  and\ \citenamefont {Stranks}}]{BRENES2017}%
  \BibitemOpen
  \bibfield  {author} {\bibinfo {author} {\bibfnamefont {R.}~\bibnamefont
  {Brenes}}, \bibinfo {author} {\bibfnamefont {D.}~\bibnamefont {Guo}},
  \bibinfo {author} {\bibfnamefont {A.}~\bibnamefont {Osherov}}, \bibinfo
  {author} {\bibfnamefont {N.~K.}\ \bibnamefont {Noel}}, \bibinfo {author}
  {\bibfnamefont {C.}~\bibnamefont {Eames}}, \bibinfo {author} {\bibfnamefont
  {E.~M.}\ \bibnamefont {Hutter}}, \bibinfo {author} {\bibfnamefont {S.~K.}\
  \bibnamefont {Pathak}}, \bibinfo {author} {\bibfnamefont {F.}~\bibnamefont
  {Niroui}}, \bibinfo {author} {\bibfnamefont {R.~H.}\ \bibnamefont {Friend}},
  \bibinfo {author} {\bibfnamefont {M.~S.}\ \bibnamefont {Islam}}, \bibinfo
  {author} {\bibfnamefont {H.~J.}\ \bibnamefont {Snaith}}, \bibinfo {author}
  {\bibfnamefont {V.}~\bibnamefont {Bulović}}, \bibinfo {author}
  {\bibfnamefont {T.~J.}\ \bibnamefont {Savenije}},\ and\ \bibinfo {author}
  {\bibfnamefont {S.~D.}\ \bibnamefont {Stranks}},\ }\bibfield  {title}
  {\bibinfo {title} {Metal halide perovskite polycrystalline films exhibiting
  properties of single crystals},\ }\href
  {https://doi.org/https://doi.org/10.1016/j.joule.2017.08.006} {\bibfield
  {journal} {\bibinfo  {journal} {Joule}\ }\textbf {\bibinfo {volume} {1}},\
  \bibinfo {pages} {155} (\bibinfo {year} {2017})}\BibitemShut {NoStop}%
\bibitem [{\citenamefont {Huynh}\ \emph {et~al.}(2022)\citenamefont {Huynh},
  \citenamefont {Liu}, \citenamefont {Chanana}, \citenamefont {Khanal},
  \citenamefont {Sercel}, \citenamefont {Huang},\ and\ \citenamefont
  {Vardeny}}]{Huynh2022}%
  \BibitemOpen
  \bibfield  {author} {\bibinfo {author} {\bibfnamefont {U.~N.}\ \bibnamefont
  {Huynh}}, \bibinfo {author} {\bibfnamefont {Y.}~\bibnamefont {Liu}}, \bibinfo
  {author} {\bibfnamefont {A.}~\bibnamefont {Chanana}}, \bibinfo {author}
  {\bibfnamefont {D.~R.}\ \bibnamefont {Khanal}}, \bibinfo {author}
  {\bibfnamefont {P.~C.}\ \bibnamefont {Sercel}}, \bibinfo {author}
  {\bibfnamefont {J.}~\bibnamefont {Huang}},\ and\ \bibinfo {author}
  {\bibfnamefont {Z.~V.}\ \bibnamefont {Vardeny}},\ }\bibfield  {title}
  {\bibinfo {title} {Transient quantum beatings of trions in hybrid organic
  tri-iodine perovskite single crystal},\ }\href
  {https://doi.org/10.1038/s41467-022-29053-6} {\bibfield  {journal} {\bibinfo
  {journal} {Nature Communications}\ }\textbf {\bibinfo {volume} {13}},\
  \bibinfo {pages} {1428} (\bibinfo {year} {2022})}\BibitemShut {NoStop}%
\bibitem [{\citenamefont {Makarov}\ \emph {et~al.}(2016)\citenamefont
  {Makarov}, \citenamefont {Guo}, \citenamefont {Isaienko}, \citenamefont
  {Liu}, \citenamefont {Robel},\ and\ \citenamefont {Klimov}}]{Makarov2016}%
  \BibitemOpen
  \bibfield  {author} {\bibinfo {author} {\bibfnamefont {N.~S.}\ \bibnamefont
  {Makarov}}, \bibinfo {author} {\bibfnamefont {S.}~\bibnamefont {Guo}},
  \bibinfo {author} {\bibfnamefont {O.}~\bibnamefont {Isaienko}}, \bibinfo
  {author} {\bibfnamefont {W.}~\bibnamefont {Liu}}, \bibinfo {author}
  {\bibfnamefont {I.}~\bibnamefont {Robel}},\ and\ \bibinfo {author}
  {\bibfnamefont {V.~I.}\ \bibnamefont {Klimov}},\ }\bibfield  {title}
  {\bibinfo {title} {Spectral and dynamical properties of single excitons,
  biexcitons, and trions in cesium--lead-halide perovskite quantum dots},\
  }\href {https://doi.org/10.1021/acs.nanolett.5b05077} {\bibfield  {journal}
  {\bibinfo  {journal} {Nano Letters}\ }\textbf {\bibinfo {volume} {16}},\
  \bibinfo {pages} {2349} (\bibinfo {year} {2016})}\BibitemShut {NoStop}%
\bibitem [{\citenamefont {Rain{\`o}}\ \emph {et~al.}(2016)\citenamefont
  {Rain{\`o}}, \citenamefont {Nedelcu}, \citenamefont {Protesescu},
  \citenamefont {Bodnarchuk}, \citenamefont {Kovalenko}, \citenamefont
  {Mahrt},\ and\ \citenamefont {St{\"o}ferle}}]{Raino2016}%
  \BibitemOpen
  \bibfield  {author} {\bibinfo {author} {\bibfnamefont {G.}~\bibnamefont
  {Rain{\`o}}}, \bibinfo {author} {\bibfnamefont {G.}~\bibnamefont {Nedelcu}},
  \bibinfo {author} {\bibfnamefont {L.}~\bibnamefont {Protesescu}}, \bibinfo
  {author} {\bibfnamefont {M.~I.}\ \bibnamefont {Bodnarchuk}}, \bibinfo
  {author} {\bibfnamefont {M.~V.}\ \bibnamefont {Kovalenko}}, \bibinfo {author}
  {\bibfnamefont {R.~F.}\ \bibnamefont {Mahrt}},\ and\ \bibinfo {author}
  {\bibfnamefont {T.}~\bibnamefont {St{\"o}ferle}},\ }\bibfield  {title}
  {\bibinfo {title} {Single cesium lead halide perovskite nanocrystals at low
  temperature: Fast single-photon emission, reduced blinking, and exciton fine
  structure},\ }\href {https://doi.org/10.1021/acsnano.5b07328} {\bibfield
  {journal} {\bibinfo  {journal} {ACS Nano}\ }\textbf {\bibinfo {volume}
  {10}},\ \bibinfo {pages} {2485} (\bibinfo {year} {2016})}\BibitemShut
  {NoStop}%
\bibitem [{\citenamefont {Nakahara}\ \emph {et~al.}(2018)\citenamefont
  {Nakahara}, \citenamefont {Tahara}, \citenamefont {Yumoto}, \citenamefont
  {Kawawaki}, \citenamefont {Saruyama}, \citenamefont {Sato}, \citenamefont
  {Teranishi},\ and\ \citenamefont {Kanemitsu}}]{Nakahara2018}%
  \BibitemOpen
  \bibfield  {author} {\bibinfo {author} {\bibfnamefont {S.}~\bibnamefont
  {Nakahara}}, \bibinfo {author} {\bibfnamefont {H.}~\bibnamefont {Tahara}},
  \bibinfo {author} {\bibfnamefont {G.}~\bibnamefont {Yumoto}}, \bibinfo
  {author} {\bibfnamefont {T.}~\bibnamefont {Kawawaki}}, \bibinfo {author}
  {\bibfnamefont {M.}~\bibnamefont {Saruyama}}, \bibinfo {author}
  {\bibfnamefont {R.}~\bibnamefont {Sato}}, \bibinfo {author} {\bibfnamefont
  {T.}~\bibnamefont {Teranishi}},\ and\ \bibinfo {author} {\bibfnamefont
  {Y.}~\bibnamefont {Kanemitsu}},\ }\bibfield  {title} {\bibinfo {title}
  {Suppression of trion formation in cspbbr3 perovskite nanocrystals by
  postsynthetic surface modification},\ }\href
  {https://doi.org/10.1021/acs.jpcc.8b06834} {\bibfield  {journal} {\bibinfo
  {journal} {The Journal of Physical Chemistry C}\ }\textbf {\bibinfo {volume}
  {122}},\ \bibinfo {pages} {22188} (\bibinfo {year} {2018})}\BibitemShut
  {NoStop}%
\bibitem [{\citenamefont {Thouin}\ \emph {et~al.}(2018)\citenamefont {Thouin},
  \citenamefont {Neutzner}, \citenamefont {Cortecchia}, \citenamefont
  {Dragomir}, \citenamefont {Soci}, \citenamefont {Salim}, \citenamefont {Lam},
  \citenamefont {Leonelli}, \citenamefont {Petrozza}, \citenamefont {Kandada},\
  and\ \citenamefont {Silva}}]{PhysRevMaterials.2.034001}%
  \BibitemOpen
  \bibfield  {author} {\bibinfo {author} {\bibfnamefont {F.}~\bibnamefont
  {Thouin}}, \bibinfo {author} {\bibfnamefont {S.}~\bibnamefont {Neutzner}},
  \bibinfo {author} {\bibfnamefont {D.}~\bibnamefont {Cortecchia}}, \bibinfo
  {author} {\bibfnamefont {V.~A.}\ \bibnamefont {Dragomir}}, \bibinfo {author}
  {\bibfnamefont {C.}~\bibnamefont {Soci}}, \bibinfo {author} {\bibfnamefont
  {T.}~\bibnamefont {Salim}}, \bibinfo {author} {\bibfnamefont {Y.~M.}\
  \bibnamefont {Lam}}, \bibinfo {author} {\bibfnamefont {R.}~\bibnamefont
  {Leonelli}}, \bibinfo {author} {\bibfnamefont {A.}~\bibnamefont {Petrozza}},
  \bibinfo {author} {\bibfnamefont {A.~R.~S.}\ \bibnamefont {Kandada}},\ and\
  \bibinfo {author} {\bibfnamefont {C.}~\bibnamefont {Silva}},\ }\bibfield
  {title} {\bibinfo {title} {Stable biexcitons in two-dimensional metal-halide
  perovskites with strong dynamic lattice disorder},\ }\href
  {https://doi.org/10.1103/PhysRevMaterials.2.034001} {\bibfield  {journal}
  {\bibinfo  {journal} {Phys. Rev. Mater.}\ }\textbf {\bibinfo {volume} {2}},\
  \bibinfo {pages} {034001} (\bibinfo {year} {2018})}\BibitemShut {NoStop}%
\bibitem [{\citenamefont {Su}\ \emph {et~al.}(2017)\citenamefont {Su},
  \citenamefont {Diederichs}, \citenamefont {Wang}, \citenamefont {Liew},
  \citenamefont {Zhao}, \citenamefont {Liu}, \citenamefont {Xu}, \citenamefont
  {Chen},\ and\ \citenamefont {Xiong}}]{Su2017}%
  \BibitemOpen
  \bibfield  {author} {\bibinfo {author} {\bibfnamefont {R.}~\bibnamefont
  {Su}}, \bibinfo {author} {\bibfnamefont {C.}~\bibnamefont {Diederichs}},
  \bibinfo {author} {\bibfnamefont {J.}~\bibnamefont {Wang}}, \bibinfo {author}
  {\bibfnamefont {T.~C.~H.}\ \bibnamefont {Liew}}, \bibinfo {author}
  {\bibfnamefont {J.}~\bibnamefont {Zhao}}, \bibinfo {author} {\bibfnamefont
  {S.}~\bibnamefont {Liu}}, \bibinfo {author} {\bibfnamefont {W.}~\bibnamefont
  {Xu}}, \bibinfo {author} {\bibfnamefont {Z.}~\bibnamefont {Chen}},\ and\
  \bibinfo {author} {\bibfnamefont {Q.}~\bibnamefont {Xiong}},\ }\bibfield
  {title} {\bibinfo {title} {Room-temperature polariton lasing in all-inorganic
  perovskite nanoplatelets},\ }\href
  {https://doi.org/10.1021/acs.nanolett.7b01956} {\bibfield  {journal}
  {\bibinfo  {journal} {Nano Letters}\ }\textbf {\bibinfo {volume} {17}},\
  \bibinfo {pages} {3982} (\bibinfo {year} {2017})}\BibitemShut {NoStop}%
\bibitem [{\citenamefont {Booker}\ \emph {et~al.}(2018)\citenamefont {Booker},
  \citenamefont {Price}, \citenamefont {Budden}, \citenamefont {Abolins},
  \citenamefont {del Valle-Inclan~Redondo}, \citenamefont {Eyre}, \citenamefont
  {Nasrallah}, \citenamefont {Phillips}, \citenamefont {Friend}, \citenamefont
  {Deschler},\ and\ \citenamefont
  {Greenham}}]{https://doi.org/10.1002/adom.201800616}%
  \BibitemOpen
  \bibfield  {author} {\bibinfo {author} {\bibfnamefont {E.~P.}\ \bibnamefont
  {Booker}}, \bibinfo {author} {\bibfnamefont {M.~B.}\ \bibnamefont {Price}},
  \bibinfo {author} {\bibfnamefont {P.~J.}\ \bibnamefont {Budden}}, \bibinfo
  {author} {\bibfnamefont {H.}~\bibnamefont {Abolins}}, \bibinfo {author}
  {\bibfnamefont {Y.}~\bibnamefont {del Valle-Inclan~Redondo}}, \bibinfo
  {author} {\bibfnamefont {L.}~\bibnamefont {Eyre}}, \bibinfo {author}
  {\bibfnamefont {I.}~\bibnamefont {Nasrallah}}, \bibinfo {author}
  {\bibfnamefont {R.~T.}\ \bibnamefont {Phillips}}, \bibinfo {author}
  {\bibfnamefont {R.~H.}\ \bibnamefont {Friend}}, \bibinfo {author}
  {\bibfnamefont {F.}~\bibnamefont {Deschler}},\ and\ \bibinfo {author}
  {\bibfnamefont {N.~C.}\ \bibnamefont {Greenham}},\ }\bibfield  {title}
  {\bibinfo {title} {Vertical cavity biexciton lasing in 2d dodecylammonium
  lead iodide perovskites},\ }\href
  {https://doi.org/https://doi.org/10.1002/adom.201800616} {\bibfield
  {journal} {\bibinfo  {journal} {Advanced Optical Materials}\ }\textbf
  {\bibinfo {volume} {6}},\ \bibinfo {pages} {1800616} (\bibinfo {year}
  {2018})}\BibitemShut {NoStop}%
\bibitem [{\citenamefont {Li}\ \emph {et~al.}(2018)\citenamefont {Li},
  \citenamefont {Wang}, \citenamefont {Lu}, \citenamefont {Jin}, \citenamefont
  {Chen}, \citenamefont {Meng}, \citenamefont {Lian}, \citenamefont
  {Taniguchi}, \citenamefont {Watanabe}, \citenamefont {Zhang}, \citenamefont
  {Smirnov},\ and\ \citenamefont {Shi}}]{Li2018}%
  \BibitemOpen
  \bibfield  {author} {\bibinfo {author} {\bibfnamefont {Z.}~\bibnamefont
  {Li}}, \bibinfo {author} {\bibfnamefont {T.}~\bibnamefont {Wang}}, \bibinfo
  {author} {\bibfnamefont {Z.}~\bibnamefont {Lu}}, \bibinfo {author}
  {\bibfnamefont {C.}~\bibnamefont {Jin}}, \bibinfo {author} {\bibfnamefont
  {Y.}~\bibnamefont {Chen}}, \bibinfo {author} {\bibfnamefont {Y.}~\bibnamefont
  {Meng}}, \bibinfo {author} {\bibfnamefont {Z.}~\bibnamefont {Lian}}, \bibinfo
  {author} {\bibfnamefont {T.}~\bibnamefont {Taniguchi}}, \bibinfo {author}
  {\bibfnamefont {K.}~\bibnamefont {Watanabe}}, \bibinfo {author}
  {\bibfnamefont {S.}~\bibnamefont {Zhang}}, \bibinfo {author} {\bibfnamefont
  {D.}~\bibnamefont {Smirnov}},\ and\ \bibinfo {author} {\bibfnamefont {S.-F.}\
  \bibnamefont {Shi}},\ }\bibfield  {title} {\bibinfo {title} {Revealing the
  biexciton and trion-exciton complexes in bn encapsulated wse2},\ }\href@noop
  {} {\bibfield  {journal} {\bibinfo  {journal} {Nature Communications}\
  }\textbf {\bibinfo {volume} {9}},\ \bibinfo {pages} {3719} (\bibinfo {year}
  {2018})}\BibitemShut {NoStop}%
\bibitem [{\citenamefont {Van~der Donck}\ \emph {et~al.}(2018)\citenamefont
  {Van~der Donck}, \citenamefont {Zarenia},\ and\ \citenamefont
  {Peeters}}]{PhysRevB.97.195408}%
  \BibitemOpen
  \bibfield  {author} {\bibinfo {author} {\bibfnamefont {M.}~\bibnamefont
  {Van~der Donck}}, \bibinfo {author} {\bibfnamefont {M.}~\bibnamefont
  {Zarenia}},\ and\ \bibinfo {author} {\bibfnamefont {F.~M.}\ \bibnamefont
  {Peeters}},\ }\bibfield  {title} {\bibinfo {title} {Excitons, trions, and
  biexcitons in transition-metal dichalcogenides: Magnetic-field dependence},\
  }\href {https://doi.org/10.1103/PhysRevB.97.195408} {\bibfield  {journal}
  {\bibinfo  {journal} {Phys. Rev. B}\ }\textbf {\bibinfo {volume} {97}},\
  \bibinfo {pages} {195408} (\bibinfo {year} {2018})}\BibitemShut {NoStop}%
\bibitem [{\citenamefont {Zhao}\ \emph {et~al.}(2024)\citenamefont {Zhao},
  \citenamefont {Huang}, \citenamefont {Gillen}, \citenamefont {Li},
  \citenamefont {Liu}, \citenamefont {Watanabe}, \citenamefont {Taniguchi},
  \citenamefont {Maultzsch}, \citenamefont {Hone}, \citenamefont {H{\"o}gele},\
  and\ \citenamefont {Baimuratov}}]{Zhao2024}%
  \BibitemOpen
  \bibfield  {author} {\bibinfo {author} {\bibfnamefont {S.}~\bibnamefont
  {Zhao}}, \bibinfo {author} {\bibfnamefont {X.}~\bibnamefont {Huang}},
  \bibinfo {author} {\bibfnamefont {R.}~\bibnamefont {Gillen}}, \bibinfo
  {author} {\bibfnamefont {Z.}~\bibnamefont {Li}}, \bibinfo {author}
  {\bibfnamefont {S.}~\bibnamefont {Liu}}, \bibinfo {author} {\bibfnamefont
  {K.}~\bibnamefont {Watanabe}}, \bibinfo {author} {\bibfnamefont
  {T.}~\bibnamefont {Taniguchi}}, \bibinfo {author} {\bibfnamefont
  {J.}~\bibnamefont {Maultzsch}}, \bibinfo {author} {\bibfnamefont
  {J.}~\bibnamefont {Hone}}, \bibinfo {author} {\bibfnamefont {A.}~\bibnamefont
  {H{\"o}gele}},\ and\ \bibinfo {author} {\bibfnamefont {A.~S.}\ \bibnamefont
  {Baimuratov}},\ }\bibfield  {title} {\bibinfo {title} {Hybrid moir{\'e}
  excitons and trions in twisted mote2--mose2 heterobilayers},\ }\href
  {https://doi.org/10.1021/acs.nanolett.4c00541} {\bibfield  {journal}
  {\bibinfo  {journal} {Nano Letters}\ }\textbf {\bibinfo {volume} {24}},\
  \bibinfo {pages} {4917} (\bibinfo {year} {2024})}\BibitemShut {NoStop}%
\bibitem [{\citenamefont {Chen}\ \emph {et~al.}(2023)\citenamefont {Chen},
  \citenamefont {Zheng}, \citenamefont {Pei},\ and\ \citenamefont
  {Zhan}}]{Chen:23}%
  \BibitemOpen
  \bibfield  {author} {\bibinfo {author} {\bibfnamefont {W.}~\bibnamefont
  {Chen}}, \bibinfo {author} {\bibfnamefont {C.}~\bibnamefont {Zheng}},
  \bibinfo {author} {\bibfnamefont {J.}~\bibnamefont {Pei}},\ and\ \bibinfo
  {author} {\bibfnamefont {H.}~\bibnamefont {Zhan}},\ }\bibfield  {title}
  {\bibinfo {title} {External field regulation strategies for exciton dynamics
  in 2d tmds},\ }\href@noop {} {\bibfield  {journal} {\bibinfo  {journal} {Opt.
  Mater. Express}\ }\textbf {\bibinfo {volume} {13}},\ \bibinfo {pages} {1007}
  (\bibinfo {year} {2023})}\BibitemShut {NoStop}%
\end{thebibliography}
\end{document}